\documentclass[12pt]{article}
\usepackage{amsmath}
\usepackage{graphicx}
\usepackage{natbib}
\usepackage{url} % not crucial - just used below for the URL 
\usepackage{xcolor}
\usepackage{hyperref}
\usepackage{mathrsfs}
\usepackage{subfigure}
\usepackage{multirow}
\usepackage{amsfonts}
\usepackage{multirow}
\usepackage{authblk}
\begin{document}

\def\spacingset#1{\renewcommand{\baselinestretch}%
{#1}\small\normalsize} \spacingset{1}

%%%%%%%%%%%%%%%%%%%%%%%%%%%%%%%%%%%%%%%%%%%%%%%%%%%%%%%%%%%%%%%%%%%%%%%%%%%%%%

%\if0\blind
%{
%  \title{\bf Bayesian Nonparametric Mixtures of Exponential Random Graph Models for Ensembles of Networks}
%  \author{Sa Ren\thanks{
%    The first author gratefully acknowledges the Graduate Teaching Assistant scholarship from University of Kent; corresponding author: Sa Ren (sr685@kent.ac.uk)}\\
%    School of Mathematics, Statistics and Actuarial Science, University of Kent, United Kingdom\\
%    Xue Wang \\
%    Walsn Limited, United Kingdom\\
%   Peng Liu\\
%    School of Mathematics, Statistics and Actuarial Science, University of Kent, United Kingdom\\
%    Jian Zhang\\
%    School of Mathematics, Statistics and Actuarial Science, University of Kent, United Kingdom\\
%}
%  \maketitle
%} \fi
%
%\if1\blind
%{
%  \bigskip
%  \bigskip
%  \bigskip
%  \begin{center}
%    {\LARGE\bf Title}
%\end{center}
%  \medskip
%} \fi

%%Note2 to Sarah, use this to Arxiv
\title{\bf Bayesian Nonparametric Mixtures of Exponential Random Graph Models for Ensembles of Networks}
\author[1]{Sa Ren\thanks{ corresponding author: Sa Ren (sr685@kent.ac.uk)}}
\author[2]{Xue Wang}
\author[1]{Peng Liu}
\author[1]{Jian Zhang}
\affil[1]{School of Mathematics, Statistics and Actuarial Science, University of Kent, Canterbury, United Kingdom, CT2 7FS}
\affil[2]{Walsn Limited, Canterbury, United Kingdom}
\maketitle

\bigskip
\begin{abstract}
Ensembles of networks arise in various fields where multiple independent networks are observed on the same set of nodes, for example, a collection of brain networks constructed on the same brain regions for different individuals. However, there are few models that describe both the variations and characteristics of networks in an ensemble at the same time. In this paper, we propose to model the ensemble of networks using a Dirichlet Process Mixture of Exponential Random Graph Models (DPM-ERGMs), which divides the ensemble into different clusters and models each cluster of networks using a separate Exponential Random Graph Model (ERGM). By employing a Dirichlet process mixture, the number of clusters can be determined automatically and changed adaptively with the data provided. Moreover, in order to perform full Bayesian inference for DPM-ERGMs, we employ the intermediate importance sampling technique inside the Metropolis-within-slice sampling scheme, which addressed the problem of sampling from the intractable ERGMs on an infinite sample space. We also demonstrate the performance of DPM-ERGMs with both simulated and real datasets.
\end{abstract}

\noindent%
{\it Keywords:}  Dirichlet process; Importance sampling; Metropolis Hastings; Slice sampling
\vfill

\newpage
\spacingset{1} % DON'T change the spacing!
%Note1 to Sarah change this to 1.5 for JCGS
\section{Introduction}\label{sec:intro}
Networks, as representations of relational data, are widely used in various scientific fields, such as sociology, neuroscience and biology.
They provide valuable insight in understanding the diverse processes behind the complex dependent interactions among different objects.
With the recent development of technology, ensembles of networks are increasingly available, which stand for multiple observations obtained on the same or similar set of nodes across different subjects or time points.
%defined on a common set of nodes(Lehmann)
Examples of ensembles of networks include a collection of brain networks from a number of participants \citep{Simpson2013}, social networks across different schools \citep{Sweet2019}, and among others.
%Comparing brain networks can provide valuable insight in diagnosing the subjects and understanding the physiological effects of certain disease or conditions.
There are high demands for developing the methodology to identify the characteristics that are common or unique across individuals by taking advantage of the wealth of data presented in an ensemble.

The statistical modeling of ensembles of networks has also been motivated by the accessibility of abundant network data.
Some researchers treat networks in an ensemble as replicates or duplicates of a true underlying network.
\cite{Durante2017} extended the latent space models using a Bayesian nonparametric approach to infer common network patterns of all networks. 
The differences across networks are ignored in this way as they assume that all networks within an ensemble have the same structure. 
In contrast, some authors argue that networks from an ensemble vary from subject to subject. %Individual differences should also be taken into account in network analysis. 
%Methods that consider the individual difference in a population including Paul and Chen (2020) \cite{Paul2020}, Arroyo et al. (2020) \cite{Arroyo2020}, Chandna and Maugis (2020) \cite{Chandna2020}, Lunag\'omez et al. (2020)\cite{Lunagomez2020}.
\cite{Paul2020} developed a random effect stochastic block model, where the individual variations from the mean community structure of the population are considered in the model. 
Similarly, \cite{Arroyo2021} introduced a common subspace independent-edge multiple random graph model that includes both the common invariant submatrix for modeling the shared latent structures and an individual score matrix for describing the individual difference.

%Chandna and Maugis (2020) \cite{Chandna2020} proposed a multi-graphon model for the non-identically distributed networks to account for both the network heterogeneity and nodal heterogeneity.
%Lunag\'omez et al. (2020)\cite{Lunagomez2020} represented multiple networks using a mean parameter (Fr\'echet mean) and a variation parameter (entropy).
%What these models have in common is that they all model the ensembles of networks from a common shared level and an individual special level. For different methods, these two levels are incorporated into the model in different ways.
Within an ensemble, some networks share common structures, while others exhibit distinct features.
Group representation is a powerful tool to capture the similarities and differences of network structures in the same ensemble.
%Incorporating group structure into network analysis is beneficial for comparing different network samples.
\cite{Durante2018} introduced a Bayesian method to test the differences between two given groups of networks.
\cite{Lehmann2021} developed a multilevel network model to compare networks from different groups.
In most cases, the underlying group structure is unknown and it is therefore necessary to develop a methodology that identifies the group membership and compares groups of networks simultaneously.
\cite{Signorelli2020} introduced a  model-based clustering method based on mixtures of generalized linear models for populations of networks.
\cite{Yin2020} proposed a finite mixture of exponential random graph models to model the ensemble of networks using the pseudo likelihood method.
However, for both models, the number of clusters need to be determined in advance. Also, the generalized linear models and the pseudo likelihood method assume the edges within a network are independent, which is not practical in real datasets.
%One advantage of ERGM is the ability to model the complex dependencies within the network. Therefore, we will conduct full Bayesian inference for our proposed model to keep the complex network structures.
%To model such samples of independent networks, we use the mixture model framework to introduce infinite mixtures of ERGMs which model multiple-networks with group-level and network-level parameters.
%The mixture model framework is beneficial for finding the underlying group structure of a population and characterize the networks with similar structures.

In this paper, we propose the Dirichlet Process Mixtures of Exponential Random Graph Models (DPM-ERGMs) for ensembles of networks. The Dirichlet process mixture model uses the Dirichlet process as a prior over an infinite mixture model, where the number of mixtures can grow adaptively with the data. This enables the model to determine the group structure of the ensemble automatically, in other words, to compare different networks without prior knowledge of the number of clusters.
Moreover, the Dirichlet process provides a large sample space and tractable posterior distributions, facilitating inference on the infinite sample space \citep{Ferguson1973}. %making it one of the most used Bayesian nonparametric models.
On the other hand, the Exponential Random Graph Model (ERGM), a versatile network model, is employed to model networks for its ability to represent various types of topological features. 
Thus, DPM-ERGMs are capable of determining the group structure and describing the group characteristics of an ensemble simultaneously.
%2.Thus, DPM-ERGMs are capable of %characterizing the ensemble by describing the number of clusters, the members of each cluster and the various connecting patterns within networks simultaneously.%or determine the group structure
%1.Thus, DPMERGMs are capable of characterizing the ensemble by flexibly clustering networks based on their topological features.
%This gives the model the flexibility to [] the ensembles with high complexity directly, avoiding the need for prior determine the structure

%difficult to know the number 
%more appropriate to have a flexible number

%The flexibility on the number of clusters allows the model to 
%It is difficult to know the number of clusters in some applications and a flexible number of clusters is required in some other situations. 

%Also, we perform full Bayesian inference to capture the dependent structures.
There are two challenges in performing Bayesian inference for DPM-ERGMs: the infinite sample space of Dirichlet process mixtures and the intractability of ERGM likelihood. 
%We develop a Metropolis-within-slice sampling algorithm to sample from the posterior distribution of DPM-ERGMs.
The slice sampling algorithm \citep{WalkerSlice2007} provides a way to sample from the posterior distribution of Dirichlet process mixture models.
%Metropolis Hastings algorithms is used inside the slice sampling algorithm to sample from the posterior distribution of ERGMs.
To sample from the infinite sample space, we borrow the idea of slice sampling and introduce a latent variable for the model, which helps us to find a finite set of components required to produce the correct Markov chain. Then the inference can be performed by sampling from the full conditional distributions of all variables on a finite space. However, the slice sampling algorithm was designed for the Dirichlet process mixtures of normal distributions, where the sampling methods related to the normal distribution are widely available. In DPM-ERGMs, sampling from posterior distributions of ERGM parameters and membership variables is challenging due to the intractable ERGM likelihood.

%To solve this problem/
%convert the infinite problem to a finite one
%determine a finite set of variables from an infinite set that are sufficient for the Markov chain.
%find the sufficient number of components that are needed to produce a valid Markov chain with the correct stationary distribution.
%Then the other problem is with the intractablity.

One way to sample from the posterior distributions of ERGMs is to use Metropolis Hastings algorithms. Standard Metropolis Hastings algorithms are not applicable since the acceptance probability depends on the intractable normalizing constants.
%Metropolis Hastings algorithm is a powerful tool to sample from the target distribution.
To address this issue, \cite{Caimo2011} applied the exchange algorithm \citep{Murray2006}, where a perfect sampler is employed to facilitate the Metropolis Hastings algorithm, avoiding the calculation of the intractable normalizing constant. 
As the perfect sampler from the ERGM is unavailable in most cases, a sample from the MCMC method is used in practice. 
%One requirement for the exchange algorithm is a perfect sampler from the ERGM, which is difficult and expensive to obtain. 
%Therefore, the quality of the sampler is important to the accuracy of the algorithm. But as Liang and Jin (2013) \cite{Liang2013} pointed out, the perfect sampler from ERGM is not practical. 
\cite{Liang2013} developed a Monte Carlo Metropolis Hastings (MCMH) algorithm to sample from the intractable posterior distributions. The algorithm is implemented by approximating the unknown normalizing constant ratio in the acceptance probability using a Monte Carlo estimate and is proved to converge to the desired target distribution. The exchange algorithm can be seen as a special case of the MCMH algorithm.
However, most of the literature on ERGMs only deals with the single network situation. 
In DPM-ERGMs, networks from the same group are multiple samples from the same ERGM distribution. This requires the Bayesian inference to have the ability of incorporating multiple network samples.

%
%M-w-S
%Importance sampling for MH
%Importance sampling for z
%IIMS
%benefit
To sample from the posterior distributions of DPM-ERGMs, we 
develop a Metropolis-within-slice sampling algorithm that employs  Metropolis Hastings inside the slice sampling algorithm. 
%We develop an Intermediate Importance sampling 
%and combine it with an Intermediate Importance sampling estimator 
%with Metropolis-within-Slice sampling algorithm, 
%A IIMS algorithm is developed, where 
%An intermediate importance sampling estimator is employed to assist the Metropolis-within-slice sampling algorithm; we refer to the combined algorithm as the IIMS algorithm.
Specifically, we extend the MCMH algorithm to a Multi-network MCMH (MMCMH) algorithm in order to update the ERGM parameters that represent multiple networks from the same group. An importance sampling estimator with intermediate values is used in MMCMH to approximate the normalizing constant ratio in the acceptance probability to ensure the accuracy of the estimation. In this way, the characteristics of the whole group can be captured by pooling information across networks. Besides, posterior samples of membership variables also suffer from the intractability issue. We express the membership variable distributions in such a way that a ratio of normalizing constants is obtained, and employ an intermediate importance sampling estimator to approximate the constructed ratio. We refer to the combined algorithm as Intermediate Importance Metropolis-within-Slice (IIMS) sampling  algorithm.
The IIMS sampling algorithm allows the full Bayesian inference to be performed based on the true likelihood, and is capable of modeling complex dependency structures beyond the pairwise interactions. Moreover, we can replace the true likelihood with the pseudo likelihood function in the Metropolis-within-slice scheme to achieve a faster, approximate computation. We will illustrate both methods in detail later.

The rest of paper is organized as follows. In Section 2, we describe how the DPM-ERGMs are formulated. Section 3 provides the sampling methodology. Section 4 presents the simulation studies. We summarize the paper in Section 5.

\section{Model Formulation}
%An ensemble of networks consists of multiple network samples. Both the individual and population network characteristics are the focuses of research interests. To achieve this, we propose to model network ensemble using infinite mixtures of ERGMs.
\subsection{Exponential Random Graph Models}
ERGMs describe the generating process of networks through exponential family distributions with summary statistics showing various connecting patterns as explanatory variables.
A network with $n$ nodes is typically represented by a random adjacency matrix $Y\in \{0,1\}^{n\times n}$, where $Y_{ij}=1$ indicates an edge between nodes $i$ and $j$, and $Y_{ij}=0$ otherwise. The realization of $Y$ is denoted by $y$ while the set of all possible outcomes of $Y$ is denoted by $\mathcal{Y}$. The covariate information regarding the nodal or network attribute that affects the connections are denoted by $X \in \mathcal{X}$.
The network structures of interest are expressed using a summary statistics vector, 
$S(y,X): \mathcal{Y} \times \mathcal{X}  \rightarrow \mathbb{R}^d$. It represents the characteristics of the network, such as the number of edges, triangles, etc, which are crucial to the formation and dissolution of networks.
The general ERGM has the following form,
\begin{align}
  P(Y=y\,|\,\theta,X)
	%=\frac{q(y;\theta)}{k(\theta)}
	=\frac{\mbox{exp}\{\theta^{\top} S(y,X)\}}{k(\theta)},\label{ergm}
\end{align}
where $\theta\in \mathbb{R}^d$ is the vector of model parameters, and $S(y,X)$ is the summary statistics \citep{morris2008}. The normalizing constant ${k\left(\theta\right)}=\sum_{ y\in \mathcal{Y}}\mbox{exp}\left\{\theta^{\top}S(y,X)\right\}$ is the sum over all potential graphs in the sample space, which is usually intractable except for very small networks. Given a realization of network $y$, the aim of statistical inference is to find which value of $\theta$ provides best description for the data under ERGM framework. The intractability of the normalizing constant is a strong barrier to the estimation of ERGMs as the likelihood function can only be specified up to a parameter dependent constant.

Bayesian inference is a natural choice for ERGMs since it allows uncertainty on model parameters.
The posterior distribution of ERGMs is
\begin{align}
 f(\theta\,|\,y,X)=\frac{\pi(\theta)P(Y=y\,|\,\theta, X)}{P(Y=y\,|\,X)},\label{pd}
\end{align}
where $\pi(\theta)$ is the prior, $P(Y=y|X)=\int_{\mathcal{R}^d}\pi(\theta) P(Y=y|\theta,X)d\theta$. The standard MCMC algorithm is not suitable since the acceptance probability as shown in \eqref{sac} to move from $\theta$ to the new proposal $\theta'$ requires evaluation of the intractable constants $k(\theta)$ and $k(\theta')$ at each step of the algorithm
\begin{align}
	\frac{\pi (\theta')h(\theta|\theta')}
	{\pi (\theta)h(\theta'|\theta)} \cdot
	\frac{\mbox{exp}\{\theta'^{\top}S(y,X)\}}	{\mbox{exp}\{\theta^{\top}S(y,X)\}}\cdot\frac{k(\theta)}{k(\theta')}.\label{sac}
\end{align}
Here, $h(\cdot)$ stands for the proposal distribution. 
MCMH algorithm \citep{Liang2013} samples from the posterior ERGMs by using an importance sampling estimator to approximate ${k(\theta)}/{k(\theta')}$ in the Metropolis Hastings algorithm.
%To sample from the posterior distribution of ERGM, \cite{Liang2013} developed a Monte Carlo Metropolis Hastings (MCMH) algorithm, where a Monte Carlo method is used to approximate the unknown ratio ${k(\theta)}/{k(\theta')}$ in the Metropolis Hastings algorithm.
%Another important issue concerning ERGMs is model degeneracy, which means the model places most of its probability mass on the empty full graph. This may cause the failure of convergence when using simulation-based MLE methods. Compared with MLE methods, Bayesian inference is considered as a better choice since it allows uncertainty on model parameters.(Bayesian pseudo might have degeneracy as well.)
\subsection{Dirichlet Process Mixtures of ERGMs}
Ensembles of networks include multiple network observations. In addition to the complex structures within each network, one may also be interested in studying the variations across different networks.
Mixture models are a natural approach to describe such a population as they can detect and characterize the subpopulations that share common structures and distinguish networks that are different automatically.
In particular, the infinite mixture model is applied here because the corresponding model complexity is adjusted to the data. %, allowing more flexibility.
%For the situation where multiple networks on the same set of nodes are available, the interest is not only on the individual network characteristics but also on the population features, which is to say, information from different networks should be pooled together. This requires a joint model that can capture both the complex structures within a network and the variations across networks in an ensemble.
%The joint modelling of all networks allows the information provided from different samples to be pooled together.
%In this way, the ensemble of networks is explored and differences and similarities across individual networks can be detected as well.
%In this research, we are interested in exploring the ensemble of networks. 
%(\textcolor{red}{more detailed explanation of ensemble networks, e.g. similarity and difference. Also what information do we want to get out of this collection of networks}).  
%Instead of modelling each individual network $ Y_i$  by an ERGM with its  individual-level  parameter $\theta_i$, 
Here, we propose to model the ensemble of networks through an infinite mixture of ERGMs, each component of which represents a cluster (subpopulation) of networks that share common structures using a cluster-specific ERGM.  

An ensemble with $N$ network samples is denoted by $\{Y_i\}_{i=1}^N$, and the corresponding covariate information is $\{X_i\}_{i=1}^N$. In such an ensemble, the single network $Y_i$ is represented using an infinite mixture of ERGMs as follows
\begin{align}
	P_{w,\theta}(Y_i=y_i\,|\,X_i)=\sum_{j=1}^{\infty}w_j\frac{\mbox{exp}\{\theta_j^{\top} S(y_i,X_i)\}}{k(\theta_j)}, \label{imm}
\end{align}
where $j$ is the cluster label, $w_j$ is the mixing proportion, $\theta_j$ is the cluster specified parameter vector, $S(y_i,X_i)$ is the summary statistics of network $y_i$, and $k(\theta_j)=\sum_{ y\in \mathcal{Y}}\mbox{exp}\left\{\theta_j^{\top}S(y,X)\right\}$ is the normalizing constant. Without requiring a fixed number of clusters in advance, the infinite mixture model is able to determine the number of clusters adaptively with the data provided. %One advantage of this proposed infinite mixture model is that it is not necessary to pre define the number of components.

The likelihood of the ensemble of networks can be expressed as
\begin{align*}
	P_{w,\theta}(\{Y_i=y_i\}_{i=1}^N\,|\,\{X_i\}_{i=1}^N)=\prod_{i=1}^N\sum_{j=1}^{\infty}w_j\frac{\mbox{exp}\{\theta_j^{\top} S(y_i,X_i)\}}{k(\theta_j)}, 
\end{align*}
%\begin{align}
%	P_{w,\theta}(Y_1=y_1,\dots,Y_N=y_N|X_1,\dots,X_N)=\prod_{i=1}^N\sum_{j=1}^{\infty}w_j\frac{\mbox{exp}[\theta_j^{\top} S(y_i,X_i)]}{k(\theta_j)}. \label{imm2.2}
%\end{align}
or
\begin{align*}
	P_{\theta}(\{Y_i=y_i\}_{i=1}^N\,|\,\{X_i, Z_i=k_i\}_{i=1}^N)=\prod_{i=1}^N\frac{\mbox{exp}\{\theta_{k_i}^{\top} S(y_i,X_i)\}}{k(\theta_{k_i})}. 
\end{align*}
where $Z=(Z_1, Z_2, \dots, Z_N)$ is a latent variable to indicate the membership of each network, e.g.  $Z_i=k_i$ if $y_i$ belongs to cluster $k_i$. 
%\begin{align}
%	P_{\theta}(Y_1=y_1,\dots,Y_N=y_N|X_1,\dots,X_N,Z_1=k_1,\dots,Z_N=k_N)=\prod_{i=1}^N\frac{\mbox{exp}[\theta_{k_i}^{\top} S(y_i,X_i)]}{k(\theta_{k_i})}. \label{imm3}
%\end{align}
It is informative to consider an infinite mixture model especially when it is not appropriate to have a limit on the number of groups. However, the inference of this model is challenging because the intractable normalizing constant has to be evaluated in the infinite sample space.
%\subsection{Prior Specification}

To perform Bayesian inference on the proposed infinite mixture of ERGMs, we adopt a Dirichlet process prior $\text{DP}(\beta,\text{H})$ \citep{Ferguson1973}, which is arguably the most commonly used Bayesian nonparametric prior. Under the constructive definition, also known as the stick-breaking representation \citep{Sethuraman1994}, 
%In the Bayesian nonparametric field, Dirichlet process (Ferguson (1973) \cite{Ferguson1973}) is widely used as a prior distribution over infinite mixture models for its large sample space and tractable posteriors, which provides a natural solution to the infinite mixture problem. By using Dirichlet process, the number of distributions required can be determined automatically and change freely with the dataset. Here, we borrow the strength of Dirichlet process prior and perform Bayesian inference for our proposed infinite mixture of ERGMs. To be precise, we use Dirichlet process prior $DP(\alpha,H)$ under its stick-breaking representation, where 
the mixing proportion $w$ is constructed using a stick-breaking procedure with an auxiliary variable $v$. A sequence of independent and identically distributed auxiliary variables $v_1, v_2, \dots$ are sampled from a prior distribution $\text{Beta} (1,\beta)$, and 
the mixing proportions are set as $w_1=v_1$, $w_j=v_j\prod_{l=1}^{j-1}(1-v_l)$ (for $j>1$). The membership indicator variable $Z$ follows a multinomial distribution $\text{Mult} (w)$ with probability $w=(w_1, w_2, \dots)$. For the prior of ERGM parameter $\theta_j$, we use a multivariate Gaussian distribution $\mathcal N(\mu_0, \Sigma_0)$. %for the prior of group related parameters $\theta_1, \theta_2, \dots$.
Given the membership $Z_i=k_i$, the network $Y_i$ is modeled by an ERGM with parameter $\theta_{k_i}$. %$P(Y_i=y_i\,|\,\theta_{k_i},X_i)$.
In the remaining of this paper, we will use Dirichlet Process Mixtures of Exponential Random Graph Models (DPM-ERGMs) with the following form, 
\begin{align}  \label{DPM}
	v_j&\sim \text{Beta} (1,\beta)\nonumber\\
	w_1&=v_1,w_j=v_j\prod_{l=1}^{j-1}(1-v_l)\\ 
	z_i&|w  \sim \text{Mult} (w)\nonumber\\
	\theta_j&|\mu_0, \Sigma_0\sim \mathcal N(\mu_0, \Sigma_0) \nonumber\\
	y_i&|Z_i=k_i,\theta \sim P_{\theta_{k_i}}(Y_i=y_i\,|\,X_i) \nonumber.
\end{align}
Here, $P_{\theta_{k_i}}(Y_i=y_i\,|\,X_i)={\mbox{exp}\{\theta_{k_i}^{\top} S( y_i,X_i)\}}/{k(\theta_{k_i})}$ is the ERGM with parameter $\theta_{k_i}$.
\section{Posterior Computation}
The statistical inference for the proposed model is very challenging due to the infinite number of mixture components and the intractable ERGM likelihood. In this section, we first develop a Metropolis-within-slice sampling algorithm to address the issue of sampling from the infinite sample space of DPM-ERGMs. %Then we show the details of the algorithm that overcomes the intractability issue and provides an accurate estimation to the original model in Section \ref{IIMS}. Moreover,  and an pseudo likelihood approximation in Section \ref{PMS} separately.Based on that, we further provide details of two algorithms that 
Then, we provide details of the algorithms based on a true and pseudo likelihood approach separately.

The slice sampling algorithm (\citealp{WalkerSlice2007}; \citealp{KalliSlice2011}) provides a way to sample from the infinite mixture components. % and Metropolis-within-slice sampling algorithm uses Metropolis Hastings algorithm inside the slice sampling to sample from the conditional distributions of model parameters in turn. We propose to combine Intermediate Importance sampling with the Metropolis-within-Slice (IIMS) sampling scheme to perform the full Bayesian inference of DPM-ERGMs. By using intermediate importance sampling, the issues of sampling from the intractable posterior ERGMs with multiple networks and membership variable distributions are addressed. Alternatively, pseudo likelihood as an approximation to the intractable likelihood can be plugged in the likelihood function and thus provides a second way of estimation. We also exhibit the pseudo likelihood Metropolis-within-Slice (PMS) sampling algorithm as a fast alternative estimation method.
%The slice sampling algorithm is conducted by sampling a sufficient but finite number of variables from the full conditional posterior distribution in turn. Metropolis-Hastings algorithm is used to sample from the posterior ERGM distribution. %More details about the slice sampling algorithm can be found in Walker (2007) \cite{WalkerSlice2007} and Kalli et al. (2011) \cite{KalliSlice2011}.
%The Metropolis-within-slice sampling algorithm is employed to generate samples from the joint posterior distributions. %This section first describes the difficulties on designing the slice sampling algorithm for our proposed model, and then presents the details of the Metropolis-within-slice sampling algorithm.
%In the case of normal distribution, it is straightforward to sample from the posterior distribution of normal distribution. However, in our model, because of the existence of intractable normalising constants, it is not possible to sample from the posterior distribution of ERGM directly. So we use a Metropolis algorithm for the sampling of the posterior distribution of ERGM parameters within the slice sampling algorithm. In this way, the Metropolis-within-slice sampling algorithm is set up.
Similar to the slice sampling, we first introduce a latent variable $u$ to our proposed model to identify the exact number of components that are required to produce a valid Markov chain with the correct stationary distributions. The joint density of $(y,u)$ is written as
\begin{align*}
	P_{w,\theta}(Y=y,u\,|\,X,\xi)
	%=&\sum_{j=1}^{\infty}\frac{w_j}{\xi_j}U(u|0,\xi_j)\xi_jP_{\theta_j}(Y=y|X)\\
	%=&\sum_{j=1}^{\infty}\frac{w_j}{\xi_j}1(u<\xi_j)P_{\theta_j}(Y=y|X)\\
	=&\sum_{j:\xi_j>u}\frac{w_j}{\xi_j}P_{\theta_j}(Y=y|X).
\end{align*}
%\begin{align*}
%P_{w,\theta}(Y=y,u|X,\xi)
%=&\sum_{j=1}^{\infty}\frac{w_j}{\xi_j}U(u|0,\xi_j)\xi_jP_{\theta_j}(Y=y|X)\\
%=&\sum_{j=1}^{\infty}\frac{w_j}{\xi_j}1(u<\xi_j)P_{\theta_j}(Y=y|X)\\
%=&\sum_{j:\xi_j>u}\frac{w_j}{\xi_j}P_{\theta_j}(Y=y|X).
%\end{align*}
%\begin{align*}
%P_{w,\theta}(Y=y,u|X)=&\sum_{j=1}^{\infty}w_jU(u|0,w_j) P_{\theta_j}(Y=y|X)\\=&\sum_{j=1}^{\infty}1(u<w_j) P_{\theta_j}(Y=y|X)\\=&\sum_{j:w_j>u}P_{\theta_j}(Y=y|X).
%\end{align*}
Compared with the original density \eqref{imm}, there are only finite numbers of $j$ satisfying $w_j>u$. In other words, the inference can be performed by sampling from the finite set $\{j:\xi_j>u\}$, which simplifies the problem dramatically.
%Also, it can be proved that integrating out $u$ from the joint density leads back to the original density in \eqref{imm}, which guarantees the accuracy of the algorithm.
%Note that $\xi=(\xi_1,\xi_2,\dots)$ can be any positive sequence, which has an effect on the speed and efficiency of the algorithm.
$\xi$ is a deterministic decreasing sequence used to address the update of $u$. See \cite{KalliSlice2011} for details and choices of $\xi$. 

Furthermore, with indicator variable $Z$, the joint density can be expressed as
\begin{align*}
	P_{w,\theta}(Y=y,u,Z=k\,|\,X,\xi)=\frac{w_{k}}{\xi_{k}}1(u<\xi_{k}) P_{\theta_{k}}(Y=y|X).
\end{align*}
Hence, the likelihood for the ensemble $\{Y_i\}_{i=1}^N$ with latent variable $u$ and sequence $\xi$ is
\begin{align}
	l_{w,\theta}(\{Y_i=y_i,Z_i=k_i,u_i\}_{i=1}^N\,|\,\{X_i\}_{i=1}^N,\xi)=\prod_{i=1}^{N}\frac{w_{k_i}}{\xi_{k_i}}{\mathbf 1}(u_i<\xi_{k_i}) P_{\theta_{k_i}}(Y_i=y_i|X_i). \label{cll}
\end{align}
With the prior distribution specified in \eqref{DPM}, the full conditional distributions of all variables $(u,w,\theta,Z)$ are available. The Metropolis-within-slice sampling scheme is performed by sampling $(u,w,\theta,Z)$ from their full conditional distributions in turn. In particular, as the direct sampling from ERGMs is not possible, Metropolis Hastings algorithm is used to assist the sampling of $\theta$.

\subsection{True likelihood based IIMS Algorithm} \label{IIMS}
In order to overcome the intractability issue and perform accurate estimation to the original model, we propose to employ the intermediate importance sampling technique in the Metropolis-within-slice sampling scheme, and name this algorithm as IIMS algorithm. %It performs accurate estimation to the model based on the true likelihood.
%Note on 30/12/2021[ By using intermediate importance sampling, the issues of sampling from the intractable posterior ERGMs with multiple networks and membership variable distributions are addressed. Specifically, we propose a Multi-network Monte Carlo Metropolis Hastings (MMCMH) algorithm to sample from the intractable posterior ERGMs with multiple network samples, where an intermediate importance sampling estimator is used to approximate the acceptance ratio. Moreover, we construct a normalizing constant ratio by introducing a constant to the posterior probability, and approximate the constructed probability ratio using intermediate importance sampling estimator. ]
The sampling procedures of the true likelihood based IIMS algorithm are listed as follows.

%{\bf (1) Sampling $u$}
\textit{Step} 1. Sample $u_i$ from a uniform distribution,
\begin{align}
u_i \sim U(0,\xi_{k_i}) \quad (i=1,2,\dots,N),
\end{align}
where $k_i$ is the current allocation of network $y_i$.

%{\bf (2) Sampling $w$}
\textit{Step} 2. Sample $v_j$ from a beta posterior distribution,
\begin{align}
	v_j \sim \text{Beta} (1+a_j, \beta+b_j) \quad (j=1,2,\dots,K^*).
\end{align}
Here, $a_j=\sum_{i=1}^N  \mathbf{1}(k_i=j)$ denotes the number of networks in group $j$ and $b_j=\sum_{i=1}^N  \mathbf{1}(k_i>j)$ corresponds to the number of networks in the groups whose label are bigger than $j$.
$K^*$ denotes the current number of clusters.
%$K^*=max\{\#\{k:\xi_k>u_i\}\}$ is the current number of clusters.

Update $w_j$ with
\begin{align}
w_1=v_1, w_j=v_j\prod_{l=1}^{j-1}(1-v_l)\quad (j=2,\dots,K^*).
\end{align}

%{\bf (3) Sampling $\theta$}
\textit{Step} 3. Sample $\theta_j$ $(j=1,2,\dots,K^*)$ using the MMCMH algorithm with the following procedures,

(1) Draw $\theta_j'$ from a proposal distribution $h(\cdot|\theta_j)$.

(2) Simulate $m_2$ networks from each intermediate distribution with parameter $\theta_r^{im}$ $(r=0,1,\dots,m_1)$ individually and store the network statistics using $S(z_{r}^s)$ $(s=1,2,\dots,m_2)$, where $\theta_r^{im}$ 
$(r=1,2,\dots,m_1)$ are $m_1$ intermediate values between $\theta_0^{im}=\theta_j$ and $\theta_{m_1+1}^{im}=\theta_j'$.

(3) Estimate the normalizing constant ratio ${k(\theta'_j)}/{k(\theta_j)}$ with an intermediate importance sampling estimator
\begin{align}
	\gamma=\prod_{r=0}^{m_1}\frac{1}{m_2}\sum_{s=1}^{m_2} \mbox{exp}\{(\theta_{r+1}^{im}-\theta_{r}^{im})^{\top}S(z_{r}^{s})\}. \label{mcmh1}
\end{align}

(4) Accept $\theta_j'$ with probability
\begin{align}
\alpha=\mbox{min}\left(1, \;\frac{\pi (\theta_j')h(\theta_j|\theta_j')}
{\pi (\theta_j)h(\theta_j'|\theta_j)} \frac{\mbox{exp}\{(\theta_j'-\theta_j)^{\top} \sum_{z_i=j}S(y_i,X_i)\}} {\gamma^{\sum_i{\mathbf{1}(z_i=j)}}}\right).
\end{align}
$\pi(\theta_j)$ is the prior distribution.

\textit{Step} 4. Sample $Z_i$ from a multinomial distribution with probability proportional to a normalizing constant dependent ratio,
\begin{align}
	P(Z_i=k_i|\cdots)\propto \mathbf{1}(\xi_{k_i}>u_i)\frac{w_{k_i}}{\xi_{k_i}} \cdot \mbox{exp}\{\theta_{k_i}^{\top}S(y_i,X_i)\}\frac{k(\theta_{c})}{k(\theta_{k_i})}, \quad (i=1,2,\dots,N). \label{alc}
\end{align}
Here, ${k(\theta_c)}$ is multiplied to construct a computable normalizing constant ratio and the normalizing constant ratios for different groups ${k(\theta_c)}/{k(\theta_{k_i})}$ $(k_i=1,\dots,K^*)$ are approximated using an intermediate importance sampling estimator as in \eqref{mcmh1}.

Remark: in \textit{Step} 3, we use a MMCMH algorithm to sample $\theta_j$ from the posterior ERGMs with multiple networks. Next, we will explain how the MMCMH algorithm is developed in Section \ref{MMCMH}. Also, we will show the construction of formula \eqref{alc} in Section \ref{SampleZ}.
\subsubsection{Sample $\theta$}\label{MMCMH}
The posterior distribution of group parameter $\theta_j$ is proportional to the product of prior $\pi(\theta_{j})$ and the joint likelihood of the networks in group $j$, which is
\begin{align}
	f(\theta_j|\cdots)\propto \pi(\theta_j)\prod_{Z_i=j}\frac{\mbox{exp}\{\theta_{j}^{\top} S( y_i,X_i)\}}{k(\theta_{j})}.\label{theta}
\end{align}
Sampling from such a posterior distribution is challenging as it depends on the product of multiple intractable likelihood functions. MCMH algorithm \citep{Liang2013} was designed to sample from the posterior ERGM of a single network. Here, we extend the MCMH algorithm to a MMCMH algorithm for the multiple network case.

In MCMH algorithm, $k(\theta_j')/k(\theta_j)$ is approximated with an importance sampling estimator
\begin{align}
%\frac{k(\theta_j')}{k(\theta_j)}=E_{\theta_j}\frac{\mbox{exp}\theta_j'^{\top}S(z^s)}{\mbox{exp}\theta_j^{\top}S(z^s)}\approx
\frac{1}{m_2}\sum_{s=1}^{m_2} \mbox{exp}\{(\theta_j'-\theta_j)^{\top}S(z^s)\}, \label{sim}
\end{align}
with $z^s$ $(s=1,2\ldots,m_2)$ denoting a sequence of $m_2$ independent auxiliary networks sampled from the ERGM with parameter $\theta_j$. %The accuracy of the estimation increases with the number of auxiliary networks $m_2$.
However, the importance sampling estimate will be incorrect if $\theta_j'$ and
$\theta_j$ are not close enough\citep{neal2005}. This obstacle can be overcome by introducing intermediate distributions between 
$\theta_j'$ and $\theta_j$. Specifically, we interpolate $m_1$ values, $\theta_r^{im}$ $(r=1,2,\dots,m_1)$, so that $\theta_r^{im}$ and $\theta_{r+1}^{im}$ are close enough,  and factorize the normalizing constant ratio using intermediate values,
\begin{align}
	\frac{k(\theta_j')}{k(\theta_j)}=\prod_{r=0}^{m_1}\frac{k(\theta_{r+1}^{im})}{k(\theta_r^{im})}=
	\frac{k(\theta_1^{im})}{k(\theta_0^{im})}
	\frac{k(\theta_2^{im})}{k(\theta_1^{im})}\cdots
	\frac{k(\theta_{m_1+1}^{im})}{k(\theta_{m_1}^{im})},
\end{align}
where $\theta_0^{im}=\theta_j$ and $\theta_{m_1+1}^{im}=\theta_j'$. 
Then, each factor ${k(\theta_{r+1}^{im})}/{k(\theta_r^{im})}$ are estimated using importance sampling estimator.

Therefore, the  intermediate importance sampling estimator to $k(\theta_j')/k(\theta_j)$ is written as 
\begin{align*}
	\gamma=\prod_{r=0}^{m_1}\frac{1}{m_2}\sum_{s=1}^{m_2} \mbox{exp}\{(\theta_{r+1}^{im}-\theta_{r}^{im})^{\top}S(z_{r}^s)\}. 
\end{align*}
where $z_r^s$ $(s=1,2,\ldots,m_2)$ is a sequence of $m_2$ independent networks sampled from the ERGM with parameter $\theta_r^{im}$.

To sample from \eqref{theta} using MMCMH algorithm, we propose $\theta_j'$ from $h(\cdot|\theta_j)$, and accept $\theta_j'$ with probability
%The probability for accepting $\theta_j'$ is
%\begin{align}
%	\frac{\pi (\theta_j')h(\theta_j|\theta_j')}
%	{\pi (\theta_j)h(\theta_j'|\theta_j)} \cdot
%	\frac{\prod_{z_i=j}{\mbox{exp}[\theta_{j}^{'\top}S( y_i,X_i)]}/{k(\theta_{j}')}}
%	{\prod_{z_i=j}{\mbox{exp}[\theta_{j}^{\top}S( y_i,X_i)]}/{k(\theta_{j})}}.
%\end{align}
\begin{align}
\frac{\pi (\theta_j')h(\theta_j|\theta_j')}
	{\pi (\theta_j)h(\theta_j'|\theta_j)} \frac{\mbox{exp}\{(\theta_j'-\theta_j)^{\top} \sum_{z_i=j}S(y_i,X_i)\}} {\gamma^{\sum_i{\mathbf{1}(z_i=j)}}}.
\end{align}
With the approximation to the normalizing constant ratio available, the acceptance ratio is calculable and thus the posterior sampling is feasible. Compared with importance sampling, the use of intermediate values increases the quality of estimation by introducing intermediate distributions. Similar techniques like annealed importance sampling and linked importance sampling \citep{neal2005} can be used as well.% Larger values for $m_1$ and $m_2$ provide better approximation but also increase computation time. %We call this adaption multi-network MCMH to distinguish with the singe-network MCMH \citep{Liang2013}.

\subsubsection{Sample $Z$} \label{SampleZ}
The full conditional distribution of $Z_i$ is
\begin{align}
P(Z_i=k_i\,|\cdots)\propto \mathbf{1}(\xi_{k_i}>u_i)\frac{w_{k_i}}{\xi_{k_i}} \cdot \frac{\exp\{\theta_{k_i}^{\top}S(y_i,X_i)\}}{k(\theta_{k_i})}.\label{z}
\end{align}
The ratio on the right hand side depends on an intractable normalizing constant $k(\theta_{k_i})$, which makes the direct sampling infeasible.
%Clearly, the values $Z_i$ can possibly take have to satisfy the condition $\{\xi_{k_i}>u_i\}$. It is enough to proceed with the chain by only looking at the set of variables that satisfy this condition. If we can identify the maximum number of clusters needed at each iteration and only perform inference on these clusters, the procedures will be simplified. Let $K_i$ be the number of components that satisfies $\{j:\xi_j>u_i\}$. Then $K=max \{K_i\}_{i=1}^N$ is the identified maximum number of clusters at each iteration. However, we also notice that the posterior membership probability depends on the intractable $k(\theta_{k_i})$, which makes the direct sampling impossible. %
%\subsubsection{Sampling from the Intractable Posterior Membership Distributions}
%The intractability also causes trouble to the sampling of membership indicator variable $Z$.
Unlike the acceptance probability, there is no normalizing constant ratio involved in the posterior membership probability. However, if we can construct a normalizing constant ratio in the posterior membership probability, we will be able to borrow the strength of intermediate importance sampling to allocate the network samples.
To do so, we multiply a constant $k(\theta_{c})$ to each term of the posterior probability vector and obtain
\begin{align*}
	P(Z_i=k_i\,|\cdots)\propto \mathbf{1}(\xi_{k_i}>u_i)\frac{w_{k_i}}{\xi_{k_i}} \cdot \mbox{exp}\{\theta_{k_i}^{\top}S(y_i,X_i)\}\frac{k(\theta_{c})}{k(\theta_{k_i})},  
\end{align*}
where the constructed normalizing constant ratios ${k(\theta_c)}/{k(\theta_{k_i})}$ $(k_i=1,2,\dots,K^*)$ using intermediate importance sampling estimation as shown in \eqref{mcmh1}.
Thus, the posterior probability ratios will not change and sampling can be performed.
%Instead, the network samples are allocated according to

%This multiplication creates a normalising constant ratio in the allocation probability and makes the calculation feasible.
%The problem left is about how to approximate $\{\frac{k(\theta_{c})}{k(\theta_1)},\cdots,\frac{k(\theta_{c})}{k(\theta_{K^*})}\}$. Importance sampling technique provides a solution to the estimation.

%Specifically, we first estimate the constructed normalising constant ratio for different groups ${k(\theta_c)}/{k(\theta_j)}$, $j=1,\cdots,K^*$, using importance sampling estimator
%with intermediate values as in \eqref{isnr}. Then we plug in the estimator to the  constructed ratio \eqref{alc} and allocate network samples based on the approximated posterior ratio. 
%The normalizing constant ratios ${k(\theta_c)}/{k(\theta_{k_i})}$ $(k_i=1,2,\dots,K^*)$ are approximated using intermediate importance sampling estimation.
The choice of $\theta_c$ is important to the accuracy of the intermediate importance sampling estimation. The estimation will be incorrect if the parameters to be compared, $\theta_c$ and $\theta_j$, are not close enough. As each group has a unique $\theta_j$, it is impossible to find one $\theta_c$ close to all $\theta_j$ at the same time. Simple importance sampling is not applicable here and multiple intermediate values must be used to ensure the quality of estimation.% Besides, parameter values near degenerate region should also be avoided because simulating network samples from ERGMs with parameters near degenerate region is troublesome. %The idea of constructing normalising constant ratio can also be used in other situations, like model selection or likelihood ratio test.

\subsection{Pseudo likelihood based PMS Algorithm}\label{PMS}
In addition to the true likelihood approach in \ref{IIMS}, we also propose a fast estimation method based on the pseudo likelihood \citep{StraussandIkeda}, which is an approximation to the true likelihood. %A fast estimation method can be obtained for DPM-ERGMs by approximating the true likelihood function with a pseudo likelihood function. %Alternatively, pseudo likelihood as an approximation to the intractable likelihood can be plugged in the likelihood function and thus provides a second way of estimation.
To be specific, the algorithm is developed by employing a pseudo likelihood approximation in the Metropolis-within-slice sampling algorithm. We name this pseudo likelihood based algorithm as PMS algorithm. In the PMS algorithm, $(u,w)$ are sampled in the same way as in the IIMS algorithm, and $(\theta,Z)$ are updated with pseudo likelihood replacement.% $\theta$ is updated using Metropolis Hastings algorithm with pseudo likelihood function approximating the intractable likelihood function in the acceptance ratio. $Z$ is sampled by estimating the likelihood in the posterior membership probability ratio using the pseudo likelihood value.

The pseudo likelihood method approximates the true likelihood using the product of conditional probabilities of all edges in a network, 
\begin{align*}
	PL_{\theta}\left(Y=y\,|\,X\right)
	%=&\prod_{r=1}^n\prod_{r\ne s}P\left(y_{rs}\,|\, y_{-rs},X\right)\\
	=&\prod_{r\ne s}P\left(y_{rs}
	=1\,|\, y_{-rs},X\right)^{y_{rs}}\left\{1-P\left(y_{rs}=1\,|\, y_{-rs},X\right)\right\}^{1-y_{rs}},
\end{align*}
where $y_{-rs}=\{y_{kl},(k,l)\ne(r,s)\}$ denotes all the dyads of the graph excluding $y_{rs}$. Here, $y_{rs}$ is described using Bernoulli distribution with probability defined by change statistics, $\Delta S_{rs}= S(y_{rs}=1,y_{-rs},X)- S(y_{rs}=0,y_{-rs},X)$, which indicates the changes of $y_{rs}$ on the summary statistics,
$$P\left(y_{rs}=1| y_{-rs}, X\right)=\frac{\mbox{exp}(\theta^{\top}\Delta S_{rs})}{1+\mbox{exp}(\theta^{\top}\Delta S_{rs})}.$$
If we replace the true likelihood with use pseudo likelihood, then the acceptance ratio for sampling  $\theta_j$ using Metropolis Hastings algorithm is 
\begin{align}
\frac{\pi (\theta_j')h(\theta_j|\theta_j')}
{\pi (\theta_j)h(\theta_j'|\theta_j)} \cdot \frac{\prod_{z_i=j}	PL_{\theta_j'}\left(Y_i=y_i|X\right)} {\prod_{z_i=j}	PL_{\theta_j}\left(Y_i=y_i|X\right)},
\end{align}
and the posterior probability of cluster membership $Z_i$ is proportional to
\begin{align}
\mathbf{1}(\xi_{j}>u_i)\frac{w_{j}}{\xi_{j}}
\cdot PL_{\theta_j}\left(Y_i=y_i\right).
\end{align}
Thus, the sampling of $\theta, Z$ is possible with the pseudo likelihood replacement. %The computational tractability makes it an attractive option.

PMS algorithm is faster than IIMS algorithm, but it is less accurate. The major issue is that it may underestimate the endogenous network formation process, since pseudo likelihood only uses local information within a whole graph \citep{Duijn2009}. 
%However, little is understood about the statistical properties of the pseudo likelihood estimators \citep{Handcock2003}. Regarding the posterior inference, \cite{Bouranis2017} showed that the Bayesian pseudo likelihood estimators are biased and their variance can be underestimated. In addition to the intractability, another important issue with ERGM is degeneracy. 
Moreover, when the model is near-degenerate, posterior samples from pseudo likelihood method may fall into the degenerate region \citep{Caimo2011}.
% We refer to the pseudo likelihood Metropolis-within-slice sampling algorithm as PMS algorithm.
%Regarding different ways of handling the intractable likelihood issue, 

%In summary, we applied the slice sampling to the network datasets here and developed a Metropolis-within-slice scheme for the posterior inference of DPMERGMs.
%which can be combined with full likelihood and pseudo likelihood.
%Full likelihood method with importance sampling estimator and pseudo likelihood approximation are demonstrated independently to handle the intractability.
%Other Bayesian inference methods on intractable likelihood can also be adapted to fit in this framework.
\section{Empirical Results}
In this section, we illustrate the performance of the proposed DPM-ERGMs through a synthetic and a real ensemble. 
%In this section, we first show the performance of the intermediate importance sampling. Then we apply DPM-ERGMs to one synthetic network ensemble and two real network ensembles with various specifications. 
The network samples from the given ERGM distribution are generated using R package ergm \citep{ergm2008}.
\subsection{Synthetic Networks}
%\tcb{In this section, we conduct simulation studies to demonstrate the performance of the proposed methods numerically. 
%We first illustrate show the advantages of full likelihood estimators over pseudo likelihood through a toy example.}
An ensemble of $N=40$ undirected networks are generated from a mixture model with $K=2$ groups. Two statistics are used to describe the networks, the number of edges $S^1(y)=\sum_{i<j}y_{ij}$ to reflect on the network density and the number of triangles $S^2(y)=\sum_{i<j<k}y_{ij}y_{jk}y_{ik}$ to represent the transitivity. The mixing proportion is $w_{true}=(0.5,0.5)$.
The network size is $n=30$. The ERGM parameters for group 1 are selected as $\theta^1_{true}=(-3,0.9)$, which has low density and high transitivity parameter, meaning that some edges are generated because of endogenous formation process. The second group parameter is $\theta^2_{true}=(-1,0)$, representing Bernoulli networks which have independent edges.
\begin{figure*}[!htbp]{}
	\centering
	{\includegraphics[width=6.5cm]{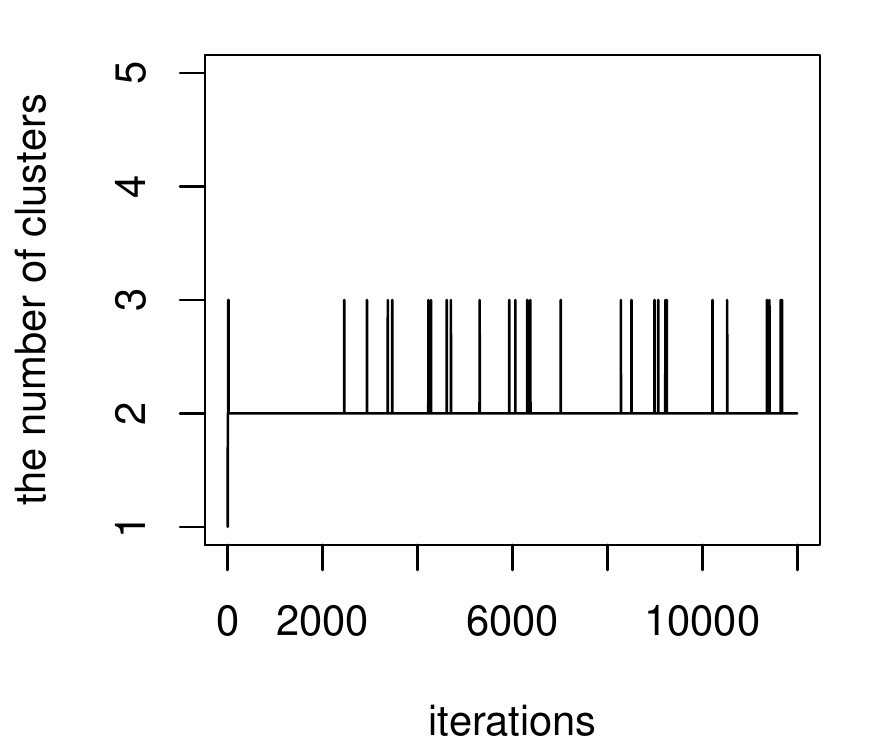}}
	{\includegraphics[width=6.5cm]{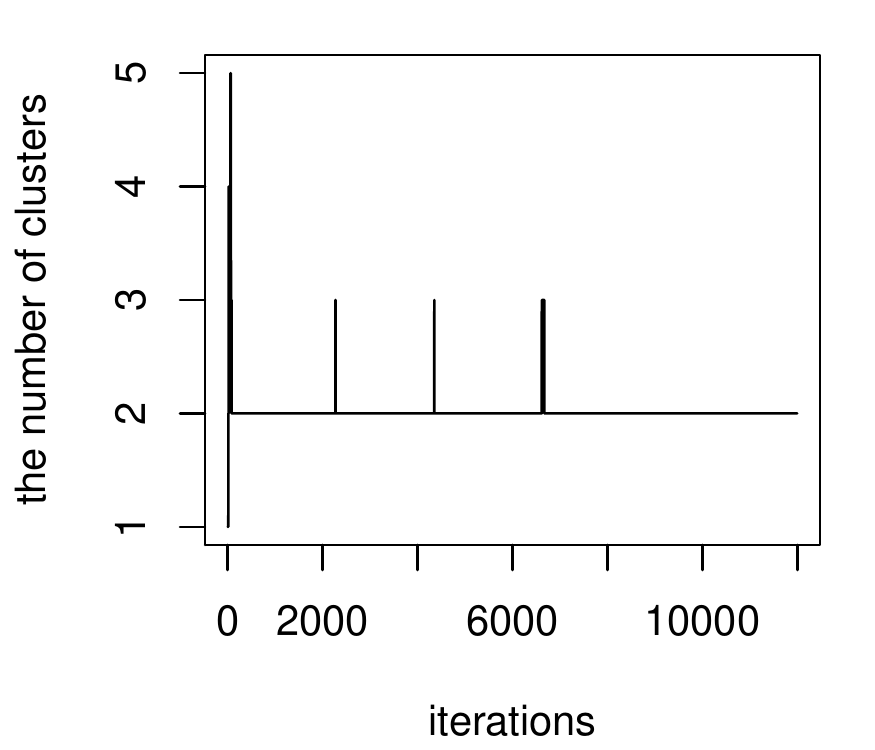}}
	%{\includegraphics[width=5.5cm]{Simulation_DIC}}
	\caption{Clustering results of IIMS algorithm (left), and PMS algorithm (right).} \label{SCF}
	%and finite pseudo likelihood method (right) for the toy example
\end{figure*}

We applied both the IIMS and PMS algorithm to the synthetic ensemble. The prior of variable $v$ is a beta distribution $\text{Beta} (1,0.1)$. The prior of ERGM parameters $\theta$ is selected to be a multivariate normal distribution $\mathcal{N} (\mu_0,\Sigma_0)$ with $\mu_0=(-3,0)$, $\Sigma_0=4^2I_2$, where $I_2$ is a two dimension diagonal matrix. %More details can be found in formula \eqref{prior}.
The proposal distribution  is $\mathcal{N}(0,\Sigma_p)$, $\Sigma_p=0.05^2I_2$. 
For sequence $\xi_1,\xi_2,\dots$, we use an exponential decreasing sequence, $\xi_i=e^{-i}$. $K_i$, the number of components that satisfies $\{j:\xi_j>u_i\}$, is also the smallest integer that satisfies $\{e^{-K_i}>u_i\}$, thus $K_i=\lfloor-log(u_i)\rfloor$.
We start with all networks in one group with initial value $\theta_0=(-2,0)$ and choose $m_1=2,m_2=10$ in the MMCMH step and $m_1=5,m_2=10$ in the sampling of membership variable. More details on the choices of $m_1, m_2$ can be found in the appendix.
\begin{figure*}[!htbp]
	\centering
	\includegraphics[width=10cm]{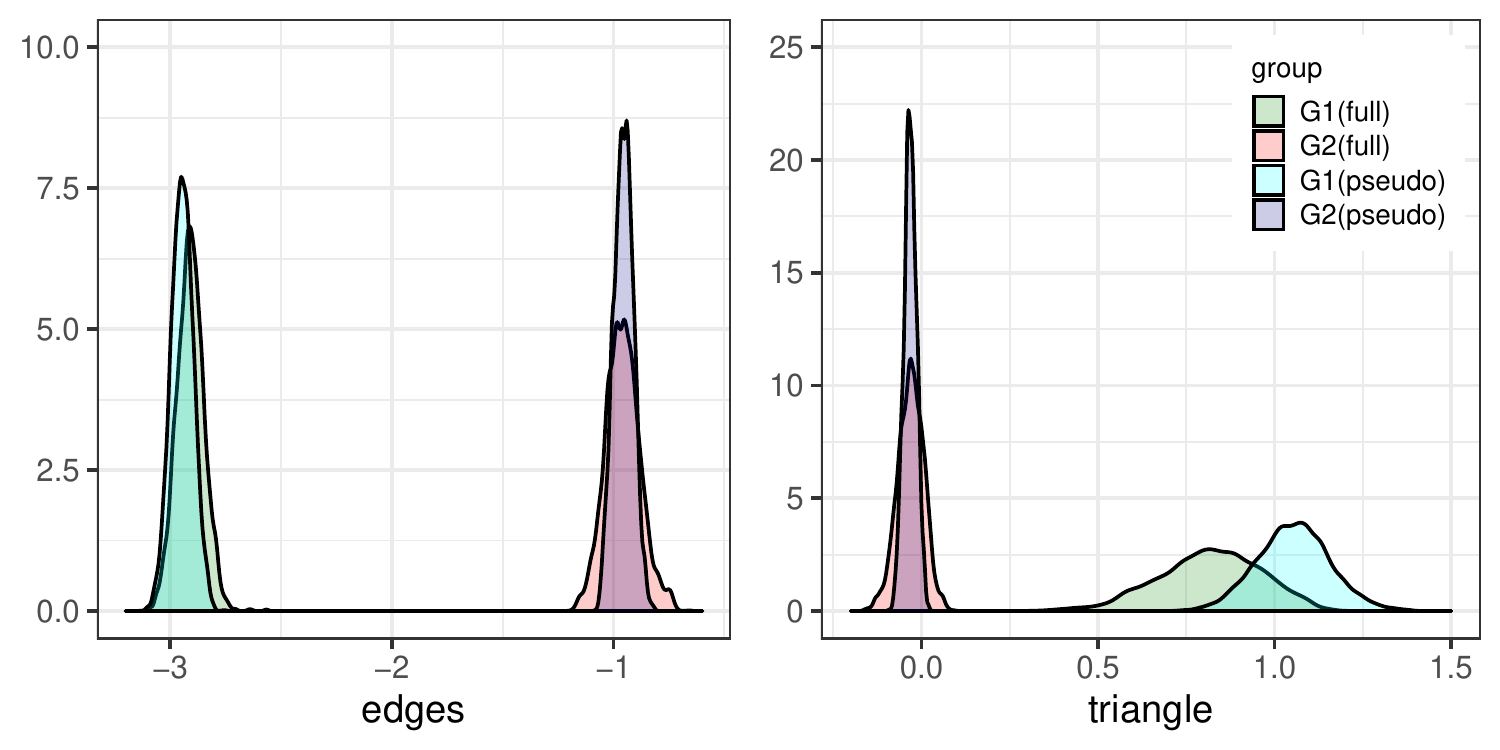}
	\caption{Density plots of ERGM parameters after 2000 burn in.} \label{SD}
\end{figure*}

The simulation is run for 12000 iterations with 2000 iterations as burn in. The clustering results are shown in Figure \ref{SCF}. Both IIMS and PMS algorithms are able to detect the true group memberships of all networks correctly. The acceptance ratio is 0.60 for group 1, 0.27 for group 2 using IIMS algorithm. The acceptance ratio is 0.62 for group 1, 0.18 for group 2 using PMS algorithm. The posterior density plots are displayed in Figure \ref{SD}. As we can see, the triangle estimator of group 1 from IIMS is smaller than PMS. This confirms the finding of \cite{Duijn2009} that pseudo likelihood method tends to underestimate the endogenous network formation process.
For Bernoulli networks in group 2, the pseudo likelihood method underestimates the parameter variance and provides a narrower interval. This is consistent with the finding of \cite{Bouranis2017}.
\begin{figure*}[!htbp]
	\centering
	\text{IIMS}\\
	\includegraphics[width=7.5cm]{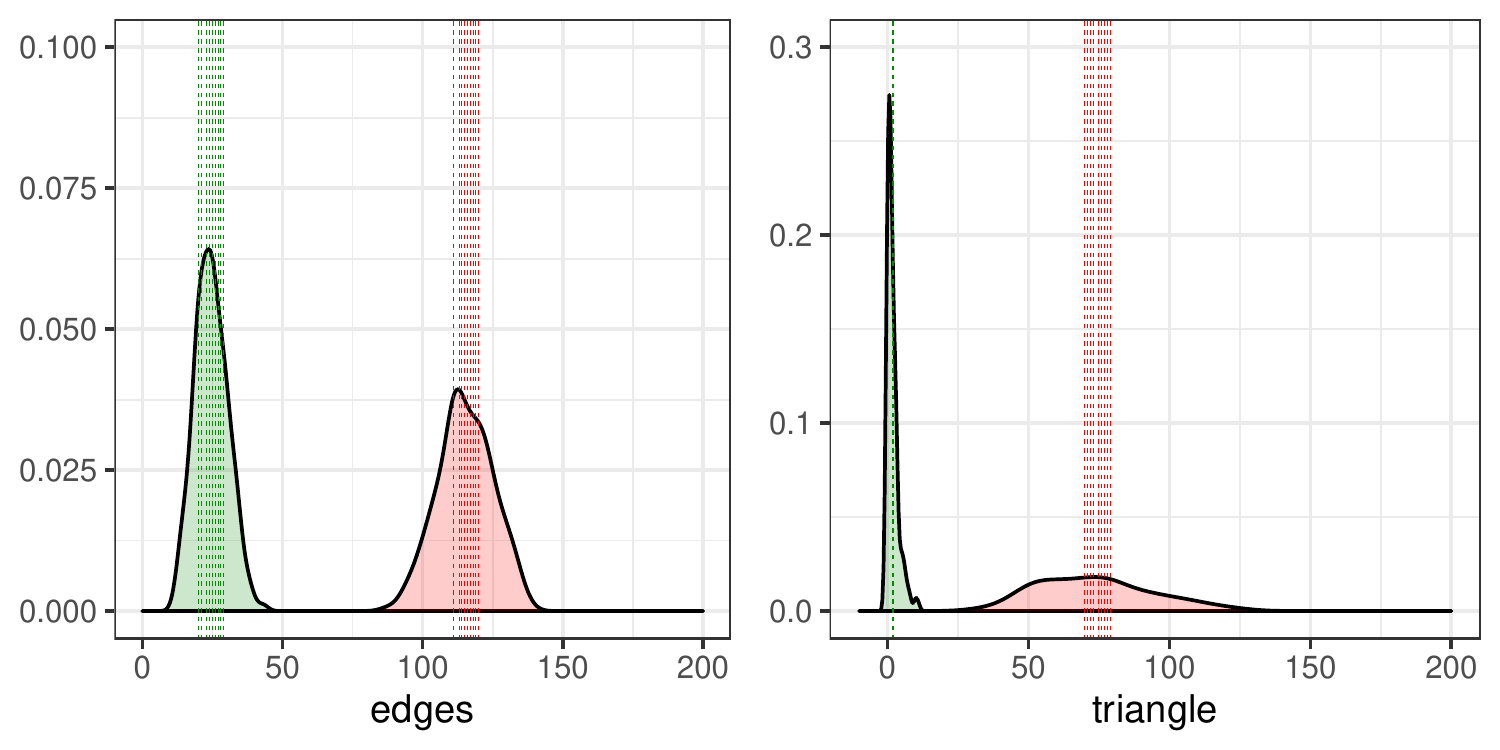}
	\includegraphics[width=7.5cm]{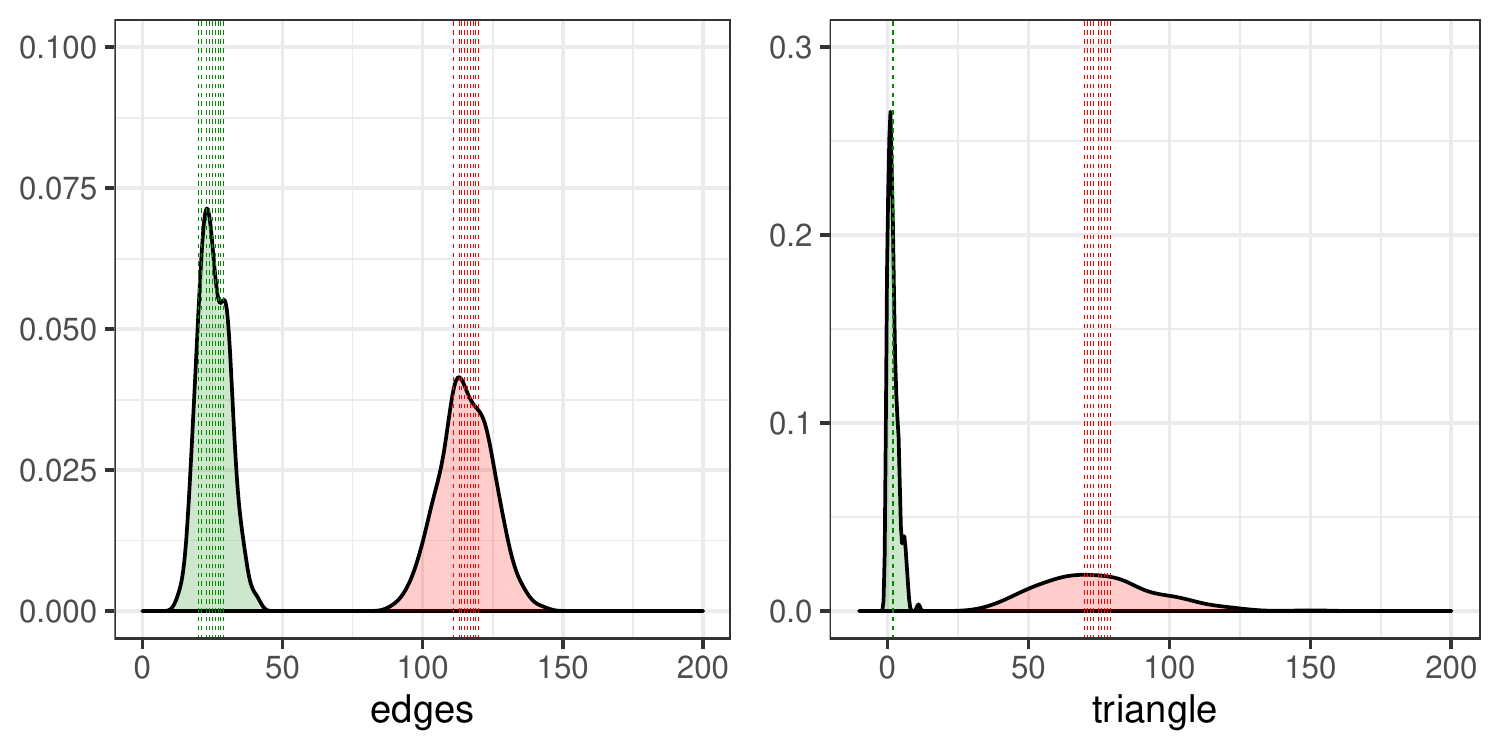}\\
	\text{PMS}\\
	\includegraphics[width=7.5cm]{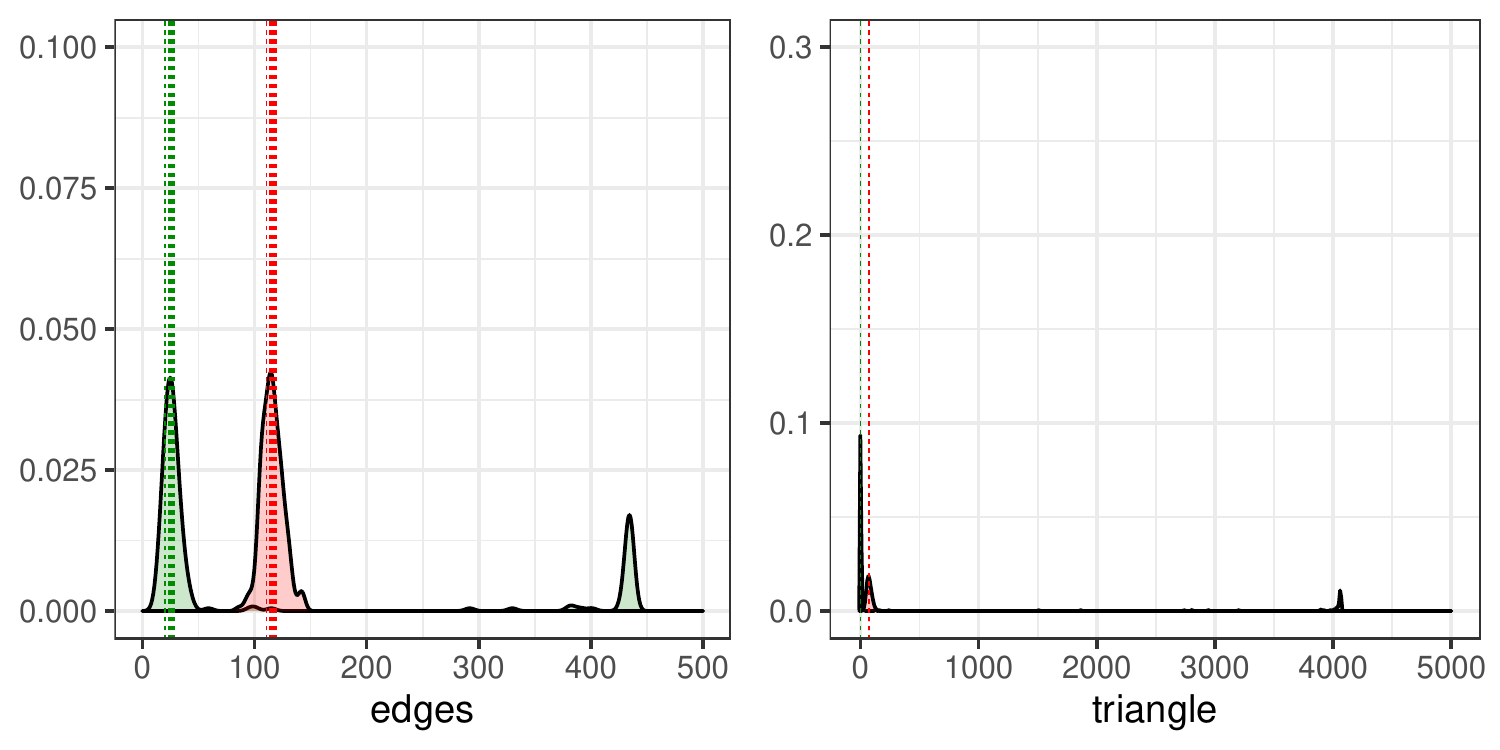}
	\includegraphics[width=7.5cm]{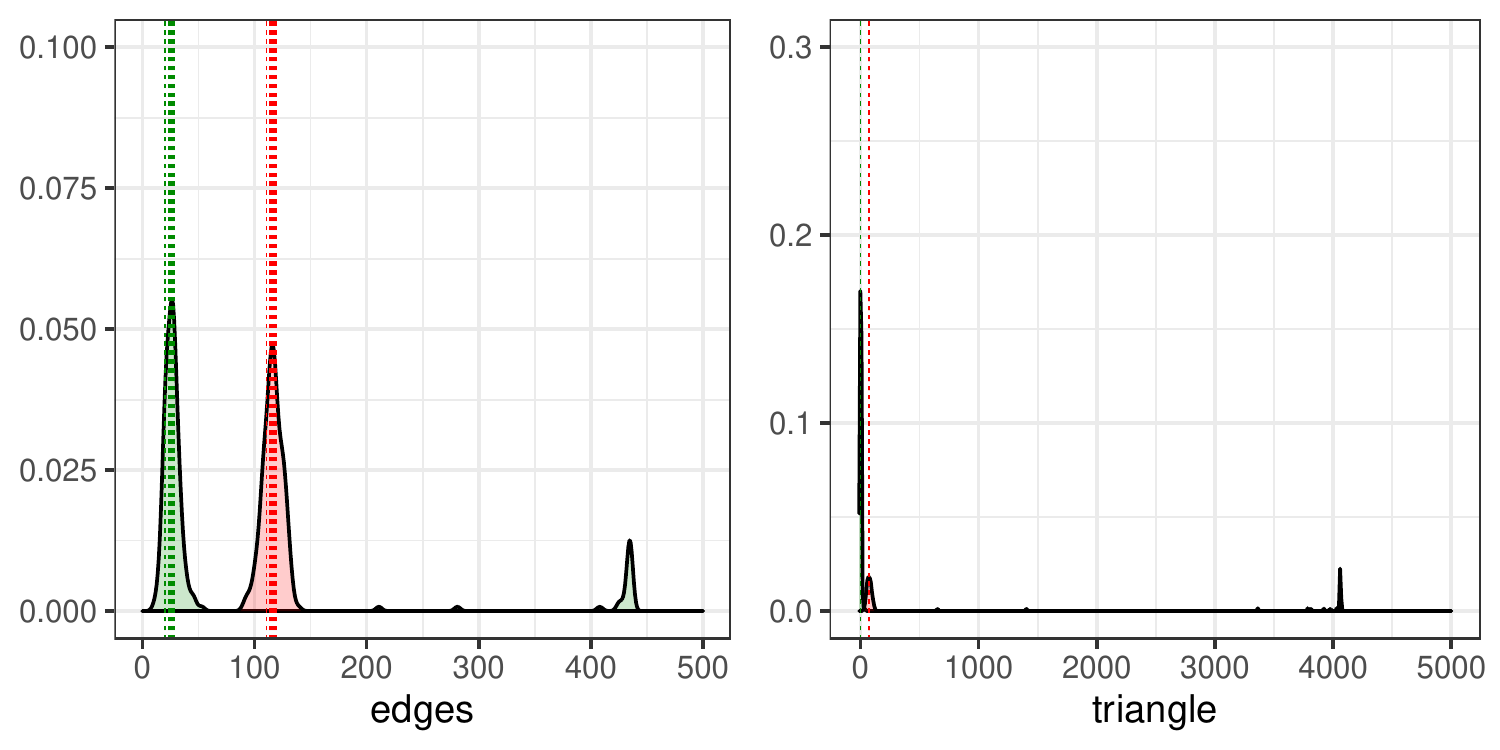}
	\caption{Density plots of simulated network statistics generated from true likelihood estimation and pseudo likelihood estimation. The left two plots are simulated from 200 different posterior samples, while the right two are sampled from posterior mean.} \label{SA}
\end{figure*}

In order to further assess the quality of estimation, we simulate networks based on the estimation. Specifically, we firstly simulate 200 networks, each from one of the 200 different posterior samples obtained after 2000 burn in and 50 thinning, then we simulate another 200 networks from the posterior mean. The simulated network statistics from true likelihood is shown in the first row of Figure \ref{SA}, with statistics from the posterior samples on the left side and statistics from the posterior mean on the right side. The observed network statistics is covered well by the simulated network statistics, indicating that the estimator is a good fit to the data. However, in the second row of Figure \ref{SA}, there are significant amount of full graphs (graphs with 435 edges) simulated from the pseudo estimation, because posterior samples from PMS method have degenerate parameter values.

In this simulation, we applied both the IIMS and PMS algorithm to the synthetic network ensemble. IIMS algorithm provided accurate estimation to the model. PMS algorithm clustered all the network samples correctly, but the estimated model for group 1 failed to generate networks resembling the observed graphs. 

\subsection{Krackhardt's Advice Networks}
%Krackhardt's High-tech Manager Networks
We next apply the proposed DPM-ERGMs to an advice network ensemble.
David Krackhardt \citep{Krack1987} studied a sequence of 21 networks about 21 employees in a high-tech machine manufacturing firm. The networks are constructed based on the data collected from a survey on the query ``Who does X go to for advice and help with work?'' Everyone is asked not only the advice relationship of themselves but also other people. Therefore, a collection of 21 perception networks $y_i (i=1,2,\dots,21)$ is built where every network represents an individual's perspective about the advice relationships among the 21 individuals. $y_{rs,i}=1$ indicates that in the opinion of individual $i, r$ asks help from $s$. The covariate information of each individual is represented by a vector $X$. 
The original paper focuses on exploring the differences of perception networks through node centrality scores to measure the importance of the nodes. Here, we are interested in learning the differences and similarities of the perception networks using the mixture of ERGMs. In this way, the generating mechanism of the perception networks can be analyzed. This helps us to better understand the perception network relationships. 
For the structure statistics, we choose the following,
%In addition to the advice relation, we also have information about each individual's age, tenure, level and department. 
\begin{itemize}
	\item $S^1(y_i)=\sum_{r\neq s}y_{rs,i}$, the total number of edges in the network. This reflects on the communication strength.
	\item $S^2(y_i)=\sum_{r\neq s}y_{rs,i}\mathbf{1}(X_r=X_s)$, the total number of connections between individuals in the same level. The positive coefficient indicates that people tend to ask for help from people of the same level, while the negative coefficient means that more help is sought  from others in a different level.
	\item $S^3(y_i)=e^{\phi}\sum_{k=1}^{n-2}\{1-(1-e^{-\phi})^k\}DP_k(y_i)$, $\phi=0.25$, geometrically weighted dyad-wise shared partner, GWDSP, a good representation for local clustering property, where $DP_k(y_i)$ represents the number of dyads with $k$ shared partners in the network $y_i$.
\end{itemize}
\begin{figure*}[!htbp]{}
	\centering
    {\includegraphics[width=6.5cm]{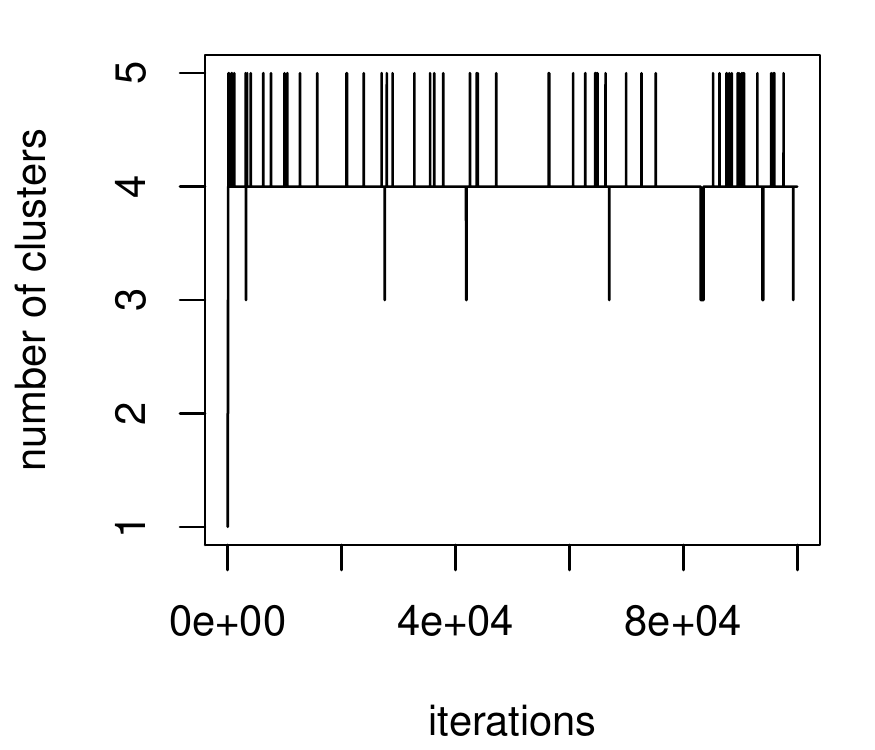}}
    {\includegraphics[width=6.5cm]{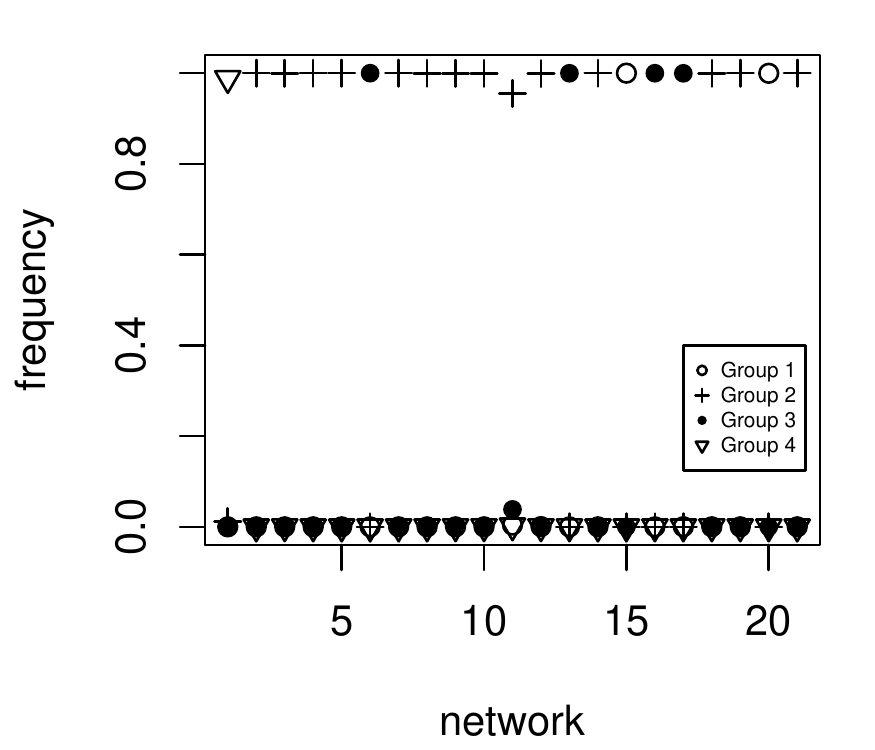}}
	\caption{Clustering results of advice network ensemble using IIMS sampling algorithm. Left: the number of clusters at every iteration. Right: the frequency of allocating to each group after 50,000 burn in.} \label{KCF}
\end{figure*}

We apply the IIMS algorithm to the advice network ensemble. The hyperparameter are specified as follows. A multivariate Gaussian distribution with mean $\mu_0=(-3,0,0)$ and covariance $\Sigma_0=4^2I_3$ is chosen as the prior distribution for ERGM parameters. The proposal variance in the MMCMH algorithm is set as $\Sigma_q=0.05^2I_3$. A beta prior $\text{Beta}(1,0.1)$ is used for the mixing proportion. $\theta_0=(-2,0,0)$ is the initial value for ERGM parameter. In the intermediate importance sampling procedure, we use $m_1=2$ intermediate distributions and $m_2=10$ auxiliary networks for MMCMH algorithm and $m_1=5, m_2=10$ in the allocation step.
\begin{figure*}[!htbp]
	\centering
	%\text{IIMS}\\
	\includegraphics[width=12cm]{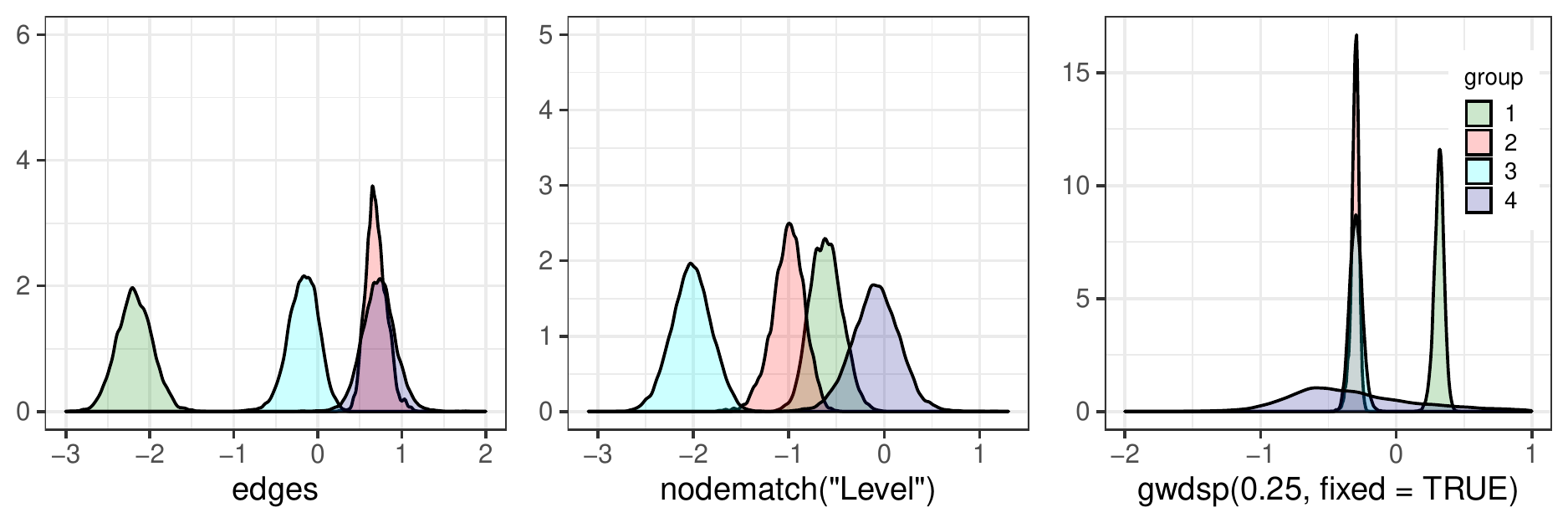}\\
	%\text{PMS}\\
	%\includegraphics[width=13cm]{Krack_pseudo_50Kburnin_density2in1_13}\\
	%\text{PMG}\\
	%\includegraphics[width=13cm]{Krack_Yin_50Kburnin_density2in1_13}
	\caption{Density plots of group parameters after 50,000 burn for advice network ensemble.% The first row is based on infinite full likelihood estimation with 4 clusters, the second row is based on infinite pseudo likelihood estimation with 6 clusters, and the third row is based on finite pseudo estimation with 4 clusters.
	} \label{KD}
\end{figure*}

The number of clusters at each iteration and the allocating frequency of each network from the IIMS algorithm are shown at Figure \ref{KCF}. We can see that 4 groups are clustered with networks 15, 20 in the first group, 2, 3, 4, 5, 7, 8, 9, 10, 11, 12, 14, 18, 19, 21 in the second group, 6, 13, 16, 17 in the third group, and network 1 in the fourth group. The acceptance probability in the MMCMH algorithm for 4 groups are 0.43, 0.16, 0.49, 0.38 respectively. 
%\begin{table}[h!]
%	\centering
%	\begin{tabular}{ |c|c|c| } 
%		\hline
%		Method & IIMS & PMS \\
%		\hline
%		Group 1 & 15, 20 & 15, 20\\ 
%		\hline
%		\multirow{2}{4em}{Group 2} &2, 3, 4, 5, 7, 8, 9,& 2, 4, 5, 8, 9, \\ 
%		&  10, 11, 12, 14, 18, 19, 21& 10, 14, 19, 21\\ 
%		\hline
%		Group 3 & 6, 13, 16, 17 & 6, 13, 16, 17\\ 
%		\hline
%		Group 4 & 1  & 1 \\ 
%		\hline
%		Group 5 & NA & 3, 7, 12, 18\\ 
%		\hline
%		Group 6 & NA  & 11\\ 
%		\hline
%	\end{tabular}
%	\caption{Clustering results.}\label{KC}
%\end{table}
%\begin{table}[h!]
%	\centering
%	\begin{tabular}{ |c|c|c|c| } 
%		\hline
%		Method & IIMS & PMS & PMG \\
%		\hline
%		Group 1 & 15, 20 & 15, 20 & 15, 20 \\ 
%		\hline
%		\multirow{2}{4em}{Group 2} &2, 3, 4, 5, 7, 8, 9,& 2, 4, 5, 8, 9, &{\bf 1}, 2, 4, 5, 8, 9,\\ 
%		&  10, 11, 12, 14, 18, 19, 21& 10, 14, 19, 21 &  10, 14, 19, 21\\ 
%		\hline
%		Group 3 & 6, 13, 16, 17 & 6, 13, 16, 17 & 6, {\bf 11}, 13, 16, 17 \\ 
%		\hline
%		Group 4 & 1  & 1 & 3, 7, 12, 18 \\ 
%		\hline
%		Group 5 & NA & 3, 7, 12, 18 & NA \\ 
%		\hline
%		Group 6 & NA  & 11 & NA \\ 
%		\hline
%	\end{tabular}
%	\caption{Clustering results.}\label{KC}
%\end{table}
To learn about the characteristics of each group, we display the posterior density plots from IIMS algorithm in Figure \ref{KD}. Group 1 has the smallest coefficient for edges but the biggest for GWDSP. This means that networks 15 and 20 have strong local clustering property, which is consistent with the fact that networks 15 and 20 have hub structures where fewer nodes have most of the connections. The advice relationships they nominate are centered around themselves. Group 2 has a big coefficient for edges and negative coefficient for level effect, indicating that networks are dense in this group and there are more advice between employees of different levels than of same levels. Group 3 has the smallest negative level effect, meaning that the advice relationships they observed are most across employees of different levels. Network 1 individually forms group 4. The level effect of network 1 is around 0, suggesting that individual level does not play a big role in network 1.

Our results are supported by the findings of \cite{Krack1987}. %, where the centrality of all nodes is calculated. 
Next, we compare our results with the centrality calculated in \cite{Krack1987}. 
Betweenness centrality reflects on the influence of a node has over the flow of information. Group 1 consists of networks 15 and 20, which have unique performances on betweenness centrality. The betweenness centrality of nodes 15, 20 is 81.15 and 65.35, which are much bigger than the rest of nodes. Both of them mentioned a lot of advice relationships they are involved in. This is consistent with our finding of local clustering phenomenon implied by high GWDSP coefficient. The networks in group 3 are distinct from the rest of individuals in terms of low indegree and betweenness centrality. The indegree of individuals 6, 13, 16, 17  is all 0, indicating that they are not asked for advice by anybody. Also, the betweenness centrality of them is 0, 0.2, 0.11, 0.28, smaller than the rest of nodes in the locally aggregated networks. Moreover, employee 1 has high indegree centrality 18, but low betweenness centrality 2.81. It is asked advice often, but rarely asks advice from other people. Of all the 18 edges individual 1 claimed, only 1 relationship is confirmed by others. The specialty of individual 1 explains why the network 1 formed a group of its own. %Network 3, 7, 12, 18 are in one group in two pseudo method results. But we haven't found explanations of this group. It is reasonable to claim that the clustering results of 
%The results of group 1 and group 3 from infinite methods are reasonable. 
%Network 1 forms a separate group in two infinite method results. Network 3, 7, 12, 18 are in one group in two pseudo method results. The ensemble of advice networks is first 

%Network 11 is also clustered to group 3 in the finite pseudo results. But we notice that the indegree centrality of node 11 is 9, much bigger than nodes 6, 13, 16, 17. It is better to not have network 11 group 3.

%\begin{figure*}[!htbp]{}
%	\centering
%	\subfigure{\includegraphics[width=13cm]{Krack_pseudo_50Kburnin_density2in1_13}}
%	\caption{Density plots of estimation for six different groups for the ensemble of advice networks. The first 50000 samples are used as burn in.}\label{TP}
%\end{figure*}
\begin{figure*}[!htbp]
	\centering
	%\text{IIMS}\\
	\includegraphics[width=13cm]{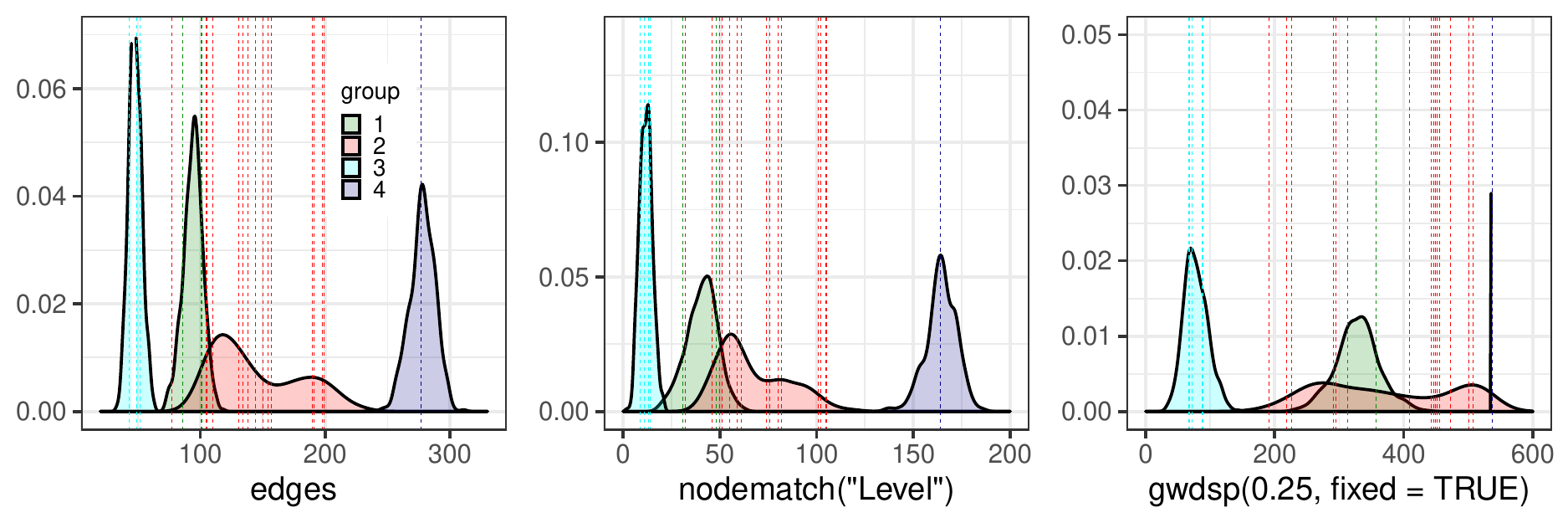}\\
	%\text{PMS}\\
	%\includegraphics[width=13cm]{921Krack_pseudo_assess1_bunin50K_thin100_sTheta_13}\\
	%\text{PMG}\\
	%\includegraphics[width=13cm]{921Krack_Yin_assess1_bunin50K_thin100_sTheta_13}
	\caption{Density plots of network statistics based on networks simulated from posterior mean. The vertical lines stand for the value of structure statistics of observed networks. %The first row is based on infinite full likelihood estimation, the second row is based on infinite pseudo likelihood estimation, and the third row is based on finite pseudo estimation with 4 clusters.
	} \label{KA}
\end{figure*}

Posterior assessments can be done by comparing the observed network statistics with simulated network statistics sampled from ERGM with estimation as parameters. Specifically, we generate 500 networks using the posterior mean as parameters and draw the density plots of the simulated network statistics in Figure \ref{KA}. As we can see, the simulated network statistics are close to the observed network statistics, suggesting that IIMS algorithm fits the data well. Note that network 1 located on the right end of the plot is far from other networks regarding the number of total edges and the number of edges within the same level. This is another reason that we think network 1 is better to be in a separate group.

\begin{figure*}[!htbp]{}
	\centering
	\includegraphics[width=6.5cm]{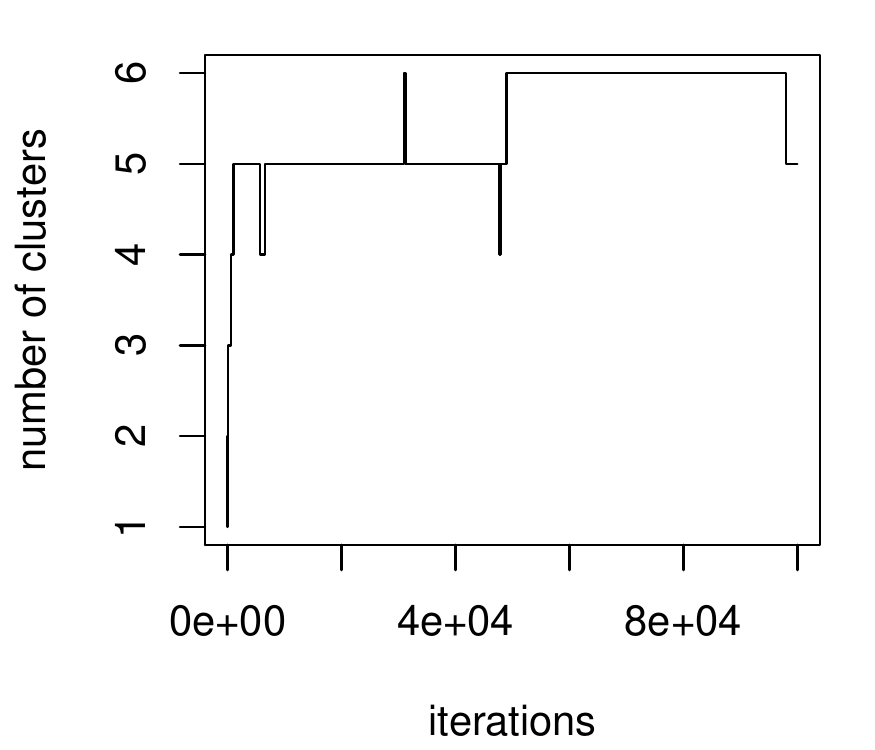}
	\includegraphics[width=6.5cm]{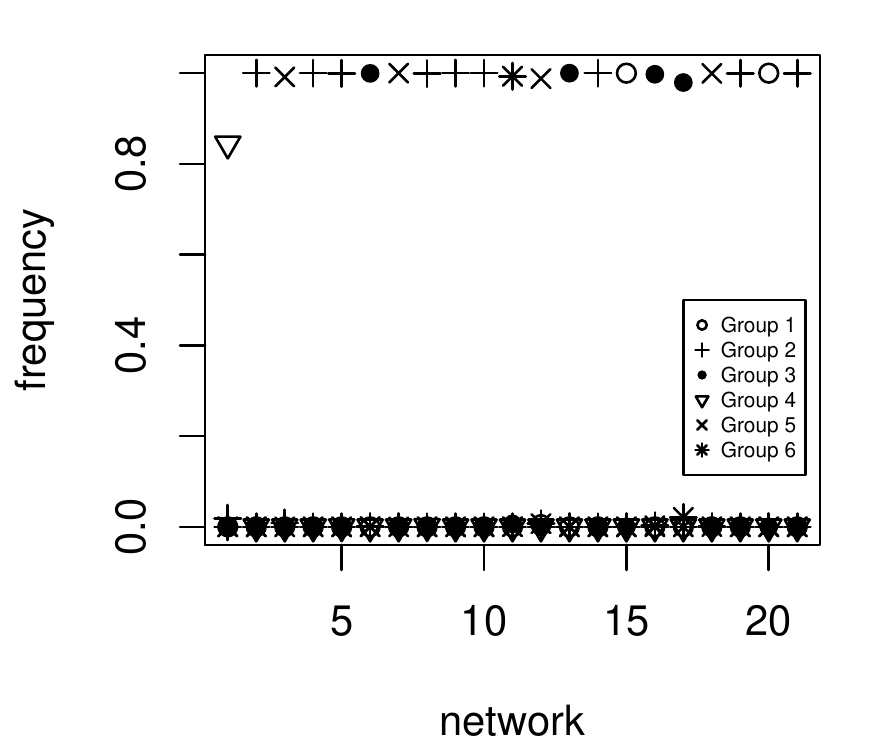}
	\caption{Clustering results of advice network ensemble using PMS sampling algorithm. Left: the number of clusters at every iteration. Right: the frequency of allocating to each group after 50,000 burn in.}\label{KCP}
\end{figure*}

Next, we apply PMS algorithm to the advice ensemble. After 100,000 iterations, 6 stable groups are detected, as shown at Figure \ref{KCP}. The networks in groups 1, 3, 4 from PMS algorithm are the same as from IIMS algorithm. The group 2 from IIMS algorithm is divided further into 3 groups, where networks 2, 4, 5, 8, 9, 10, 14, 19, 21 form the new second group, 3, 7, 12, 18 make the new fifth group, and 11 is in the sixth group. The acceptance probability of the MMCMH algorithm for 6 groups are 0.36, 0.23, 0.40, 0.36, 0.29, 0.50 respectively. 
 % \label{KCP} %this location shows label with chapter as well

\begin{figure*}[!htbp]{}
	\centering
	\includegraphics[width=12cm]{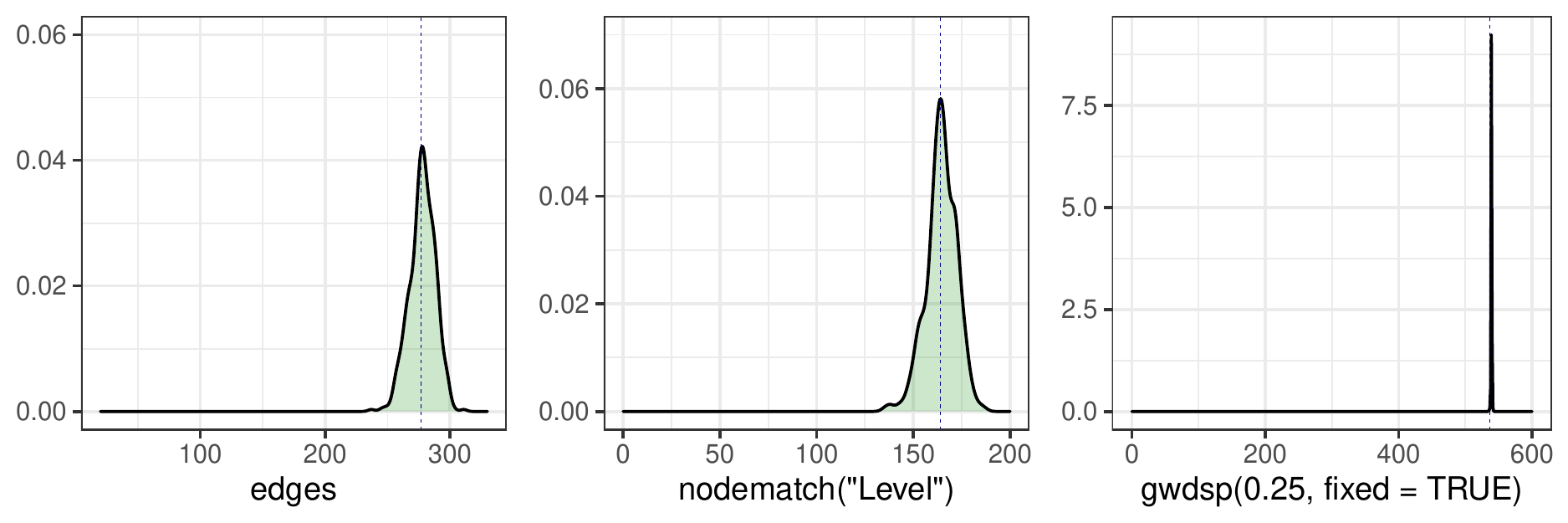}\\
	\includegraphics[width=12cm]{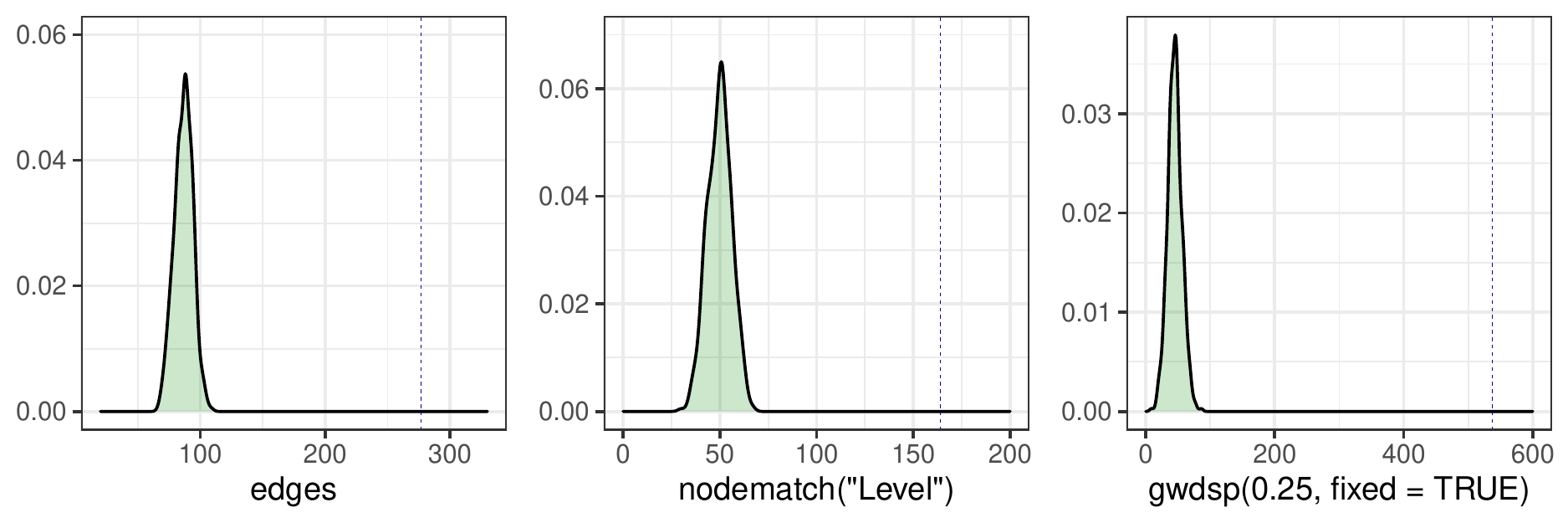}
	\caption{Simulated network statistics from IIMS estimation for network 1 (or group 4) is in the first row. Simulated network statistics from PMS estimation for network 1 (or group 4) is in the second row. }\label{Kdeg}
\end{figure*}

Furthermore, we calculate the distance between observed network statistics and simulated network statistics as follows, 
\[\sum_{i: Z_i=k}(S(y_i)-\frac{\sum_{l=1}^{500} S(z_l^k)}{500})^2 \quad (k=1,2,\dots,K),\] 
where $S(y_i)$ represents the summary statistics of observed network $y_i$, and $S(z_l^k)$ stands for the summary statistics of simulated networks from ERGM with group parameter $\theta_{k}$.
%Furthermore, as in the previous dataset, we also measure the distance between each observed network statistics to the average of 500 simulated network statistics from model estimation.
The results of are shown in Table \ref{KDis}.
\begin{table}[h!]
	\caption{The distance between observed network statistics and simulated network statistics.} \label{KDis}
	\begin{center}
	\begin{tabular}{ccccc} 
		Method & Group 1  & Group 2 & Group 3 & Group 4 \\\hline
		IIMS & 1310 & 191507 & 411 & 6\\ 
		PMS & 1358 & 134267, 10162, 32 &720 &291048\\ 
	\end{tabular}
	\end{center}
\end{table}
Comparing both results, the estimation for groups 1, 3, 4 from IIMS is more accurate, especially that the IIMS estimation of group 4 is much better than the PMS estimation. To get more details, we show the density plots of the simulated network statistics on Figure \ref{Kdeg}. Simulated network statistics from IIMS are centered around the observed statistics on the top row, while simulated statistics from PMS are distant from the observed statistics on the second row. This is because the model for group 4 is near-degenerate. For a near-degenerate model, the underlying parameter values are close to a degenerate region, which increases the difficulty for estimation. This can happen quite often when we fit a ERGM with complicated statistics to real datasets. The pseudo likelihood method does not work for the near-degenerate model \citep{Caimo2011}. In this case, we can only use true likelihood method. For the 14 networks in group 2, the total distance is smaller for PMS method. This is understandable because the IIMS method fits all these 14 networks with one model, while the PMS method fits these networks with 3 models. 
%\begin{figure*}[!htbp]{}
%	\centering
%	\includegraphics[width=13cm]{Krack_full_assess1_gro_4}\\
%	\includegraphics[width=13cm]{Krack_pseudo_assess1_gro_4}
%	\caption{Simulated network statistics from full likelihood estimation for network 1 (or group 4) is in the first row. Simulated network statistics from pseudo likelihood estimation for network 1 (or group 4) is in the second row. }
%\end{figure*}
%\begin{figure*}[!htbp]{}
%	\centering
%	\includegraphics[width=13cm]{921Krack_full_assess1_gro_4_dTheta}\\
%	\includegraphics[width=13cm]{921Krack_pseudo_assess1_gro_4_dTheta}
%	\caption{Simulated network statistics from full likelihood estimation for network 1 (or group 4) is in the first row. Simulated network statistics from pseudo likelihood estimation for network 1 (or group 4) is in the second row. }
%\end{figure*}

%\begin{figure*}[!htbp]{}
%	\centering
%	\includegraphics[width=7cm]{krack_DIC}
%	%\includegraphics[width=7cm,height=6cm]{Krack_Yin_z50K_freq}
%	\caption{DIC values for different number of clusters using PMG sampling algorithm.}\label{KCY}
%\end{figure*}

%In addition, we also applied the PMG sampling algorithm to the advice ensemble with the number of clusters $k=1,2,3,4,5,6$.
%DIC values with different number of clusters are shown in the left panel of Figure \ref{KCY}. The optimal number of clusters is the one with the lowest DIC value. However, in this application, the DIC value does not have a lowest point. To avoid overfitting, we choose the number of clusters by identifying the elbow point of the curve. One way to do so is to select the fist number that has the changing ratio bigger than -0.005. Under this rule, we choose the number of clusters to be 4 and present the allocation frequency of each network in the right panel of Figure \ref{KCY}. Next, we explain the model results by comparing with the results of \cite{Krack1987}.
In this simulation, we applied the IIMS algorithm to the advice ensemble and found 4 meaningful clusters. Although pseudo likelihood based methods managed to divide the ensemble into reasonable clusters, they failed to represent the features of networks because they are not suitable for estimating the near-degenerate model in this example.  More simulation results can also be found in the appendix.
%In summary, the clustering results of three methods are consistent. Reasonable groups are divided and confirmed. The infinite mixture model is flexible as it determines the number of clusters automatically. It is possible to fit finite mixture models with different number of clusters to the ensembles of networks and select the optimal one. However, it requires more work and there is no clear rules on deciding the number of clusters. Pseudo likelihood method can provide reasonable estimation the networks, but it can fail when the model is near-degenerate.

\section{Discussion}
In this paper, we proposed to model the ensemble of networks using a Dirichlet process mixture of ERGMs. 
Through such a framework, the subpopulations consisting of similar networks can be detected and compared automatically without requiring a fixed number of clusters in advance. On the other hand, multiple networks with similar characteristics are described by the same ERGM, namely, the cluster-specific ERGM, which is better than a single network ERGM, because information from all networks in the same cluster are gathered together on the cluster-specific ERGMs.
Moreover, we also developed a novel IIMS sampling algorithm for the full Bayesian inference of the DPM-ERGMs in order to capture the higher order interactions within a network.
%A latent variable is introduced to facilitate the sampling from the infinite space. An intermediate importance sampling estimator is utilized to approximate the intractable normalising constant ratio, which is a big problem as the ERGM likelihood with intractable normalising constant has to be evaluated both at the Metropolis-Hastings step and allocation step. 
%Such a framework can also be used to other models with intractable normalising constant.

The full Bayesian inference of ERGMs is known to be time consuming as generating networks from desired ERGMs requires a long run of Markov chain using MCMC technique. We provided a PMS sampling algorithm as a fast approximation method which can be used for pre-analysis of the dataset. However, as we mentioned before, PMS algorithm can not capture the higher order interactions within the network and can fail estimation when the model is near-degenerate. For a more accurate estimation, IIMS sampling algorithm is recommended. 
%\section*{Supplementary Materials}
\section*{Acknowledgments}
Sa Ren was supported by the Graduate Teaching Assistant scholarship from University of Kent. 
%The authors would like to thank Prof. Jian Zhang and Dr. Peng Liu for the helpful comments on the paper.
The authors report there are no competing interests to declare.
%The authors would like to thank the anonymous referees, an Associate Editor and the Editor for their constructive comments that improved the 	quality of this paper.
%\section*{Funding}
%Sa Ren was supported by the Graduate Teaching Assistant scholarship from University of Kent. 
%\section*{ORCID}
%Sa Ren https://orcid.org/0000-0001-9040-249X
\bibliographystyle{Chicago}
\bibliography{C://Users/SR685/Desktop/Writing/reference}
\section*{Appendix}
\subsection*{1.1  Intermediate Importance Sampling}
%Intermediate importance sampling technique is employed to approximate the normalizing constant ratio as expressed in \eqref{mcmh1}.
Here, we use simulation studies to show how the number of intermediate distributions $m_1$ and the number of auxiliary networks $m_2$ affect the normalizing constant ratio approximation with varying distances between compared parameters.
%The estimation will be poor if the parameters to be compared are not close enough. 
%The number of intermediate distributions $m_1$ and the number of auxiliary networks $m_2$ have to be chosen in advance.
%Larger values of $m_1$ and $m_2$ provide better estimations but also take longer time.
%The choice is also related to the the distance of parameters to be compared.
%The normalising constants in Metropolis-Hastings algorithm are closer to each other, compared with those in the sampling of posterior membership distribution. 
%Smaller values of $m_1$ and $m_2$ are enough for MCMC step. In order to select optimal values for $m_1$ and $m_2$, it is necessary to do some simulations in advance.
\begin{figure}[!htbp]
	\begin{center}
		\includegraphics[width=4.5cm]{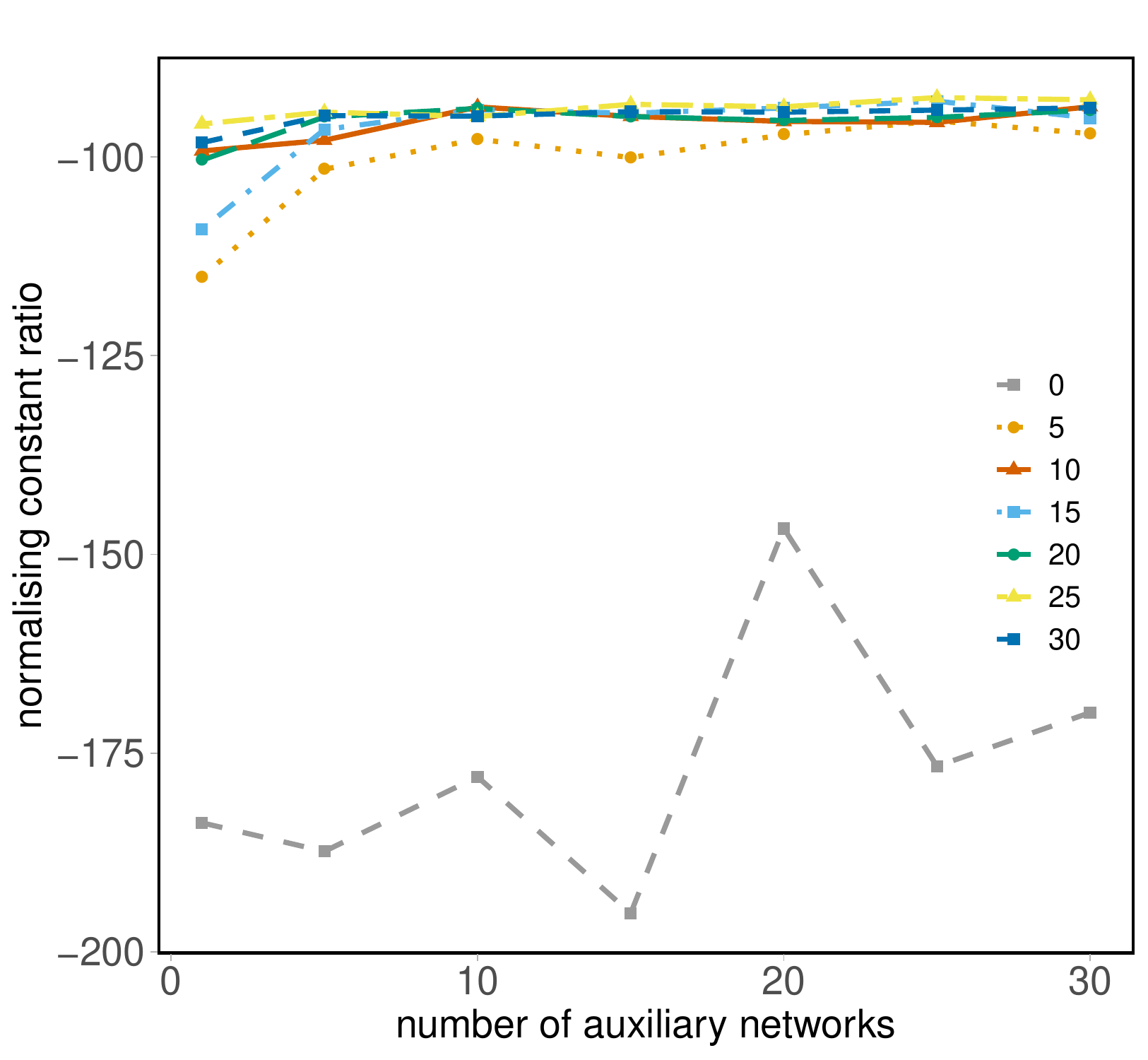}
		\includegraphics[width=4.5cm]{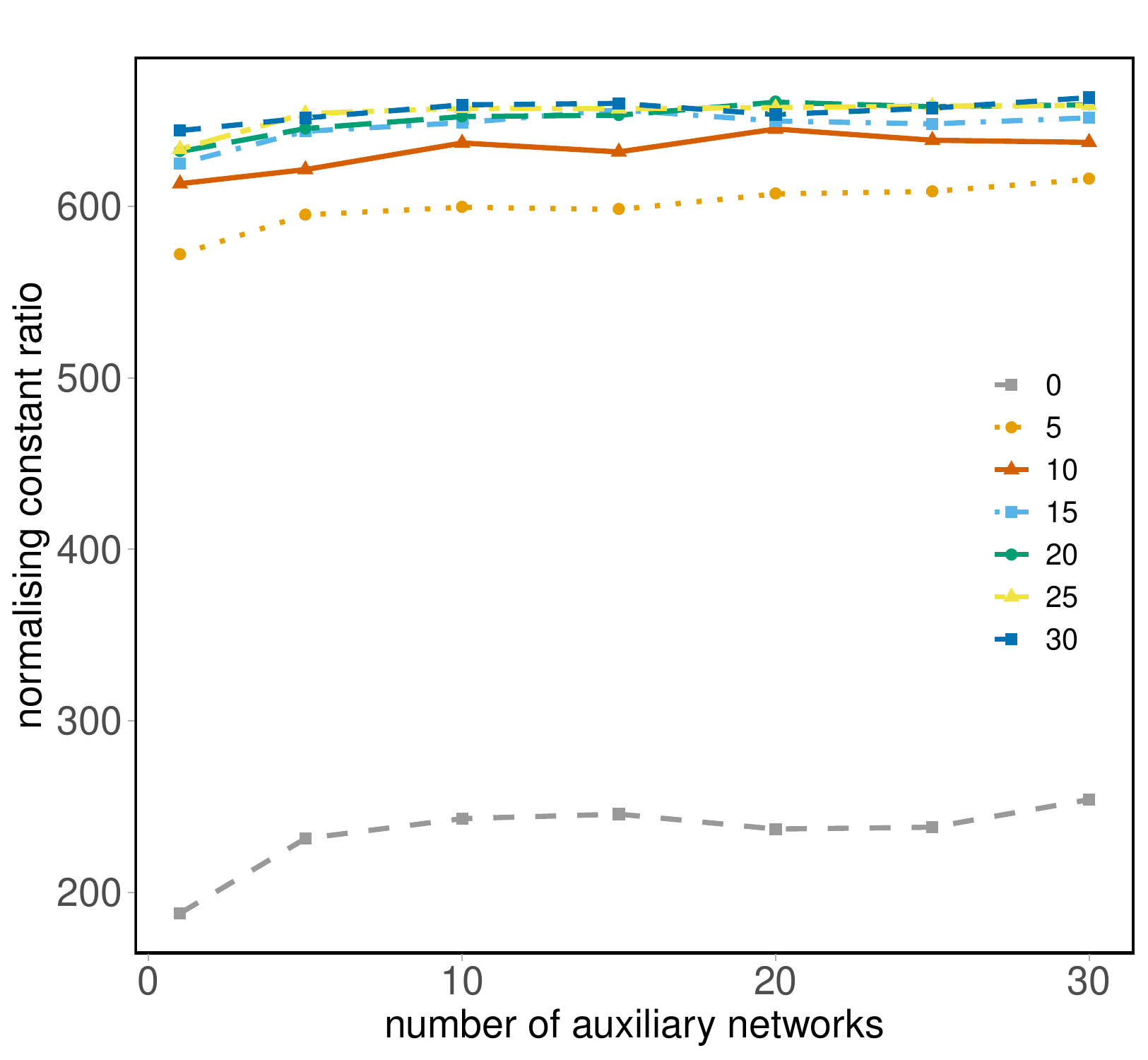}
		\includegraphics[width=4.5cm]{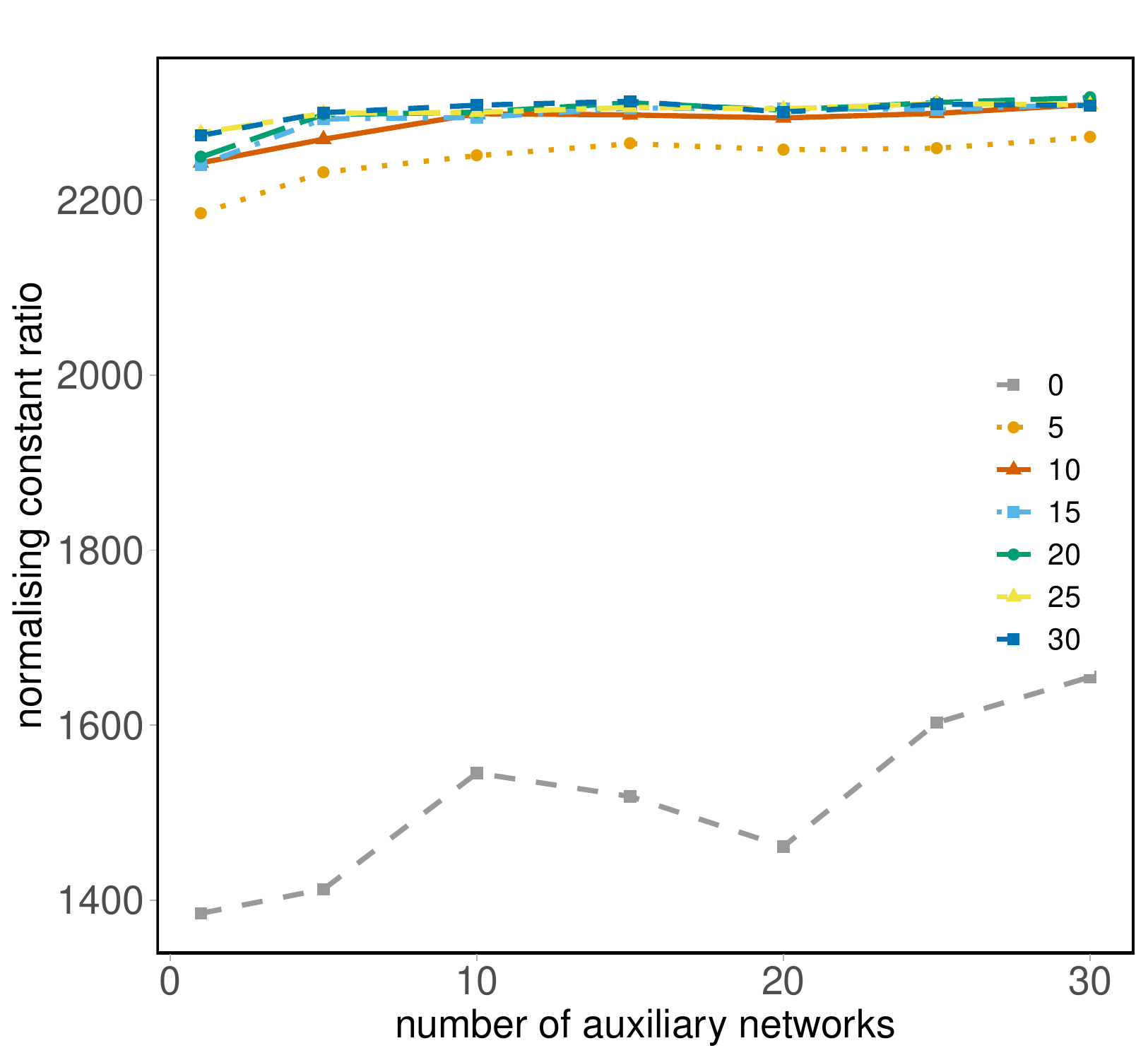}
	\end{center}
	\caption{Estimation of normalizing constant ratios when the compared parameters are far from each other. Three plots correspond to three repetitions of the parameters.}\label{nc1}
\end{figure}

We first show that how the estimation changes with different values of $m_1$ and $m_2$ when the compared parameters are distant. To do so, we sample two parameters $\theta_1, \theta_2$ independently from the prior $\mathcal{N} (\mu_0,\Sigma_0)$ and estimate $k(\theta_2)/k(\theta_1)$ with different values of $m_1, m_2$. The results of three repetitions are shown in the three plots of Figure \ref{nc1} separately. In each plot, lines with different colors correspond to different numbers of intermediate values $m_1$ and x-axis represents different numbers of auxiliary variables $m_2$. %Three plots correspond to different pairs of $(\theta_1, \theta_2)$.
As is shown, the line of $m_1=0$ is far from the other lines, meaning that the estimation is incorrect and intermediate distributions have to be used to get a good estimation. %Also, with increasing $m_1$
%We recommend $m_1=5, m_2=10$ for posterior membership sampling step.%All lines merge together with increasing $m_1$ and $m_2$, meaning that except for $m_1=0$ when there is no intermediate distributions used. %We choose $m_1=2,m_2=10$ in the MCMC step and $m_1=5,m_2=10$ in the allocation step. Here, we provide a way to initialize the values of $m_1, m_2$, which can be adjusted during iterations.
\begin{figure}[!htbp]
	\begin{center}
		\includegraphics[width=4.5cm]{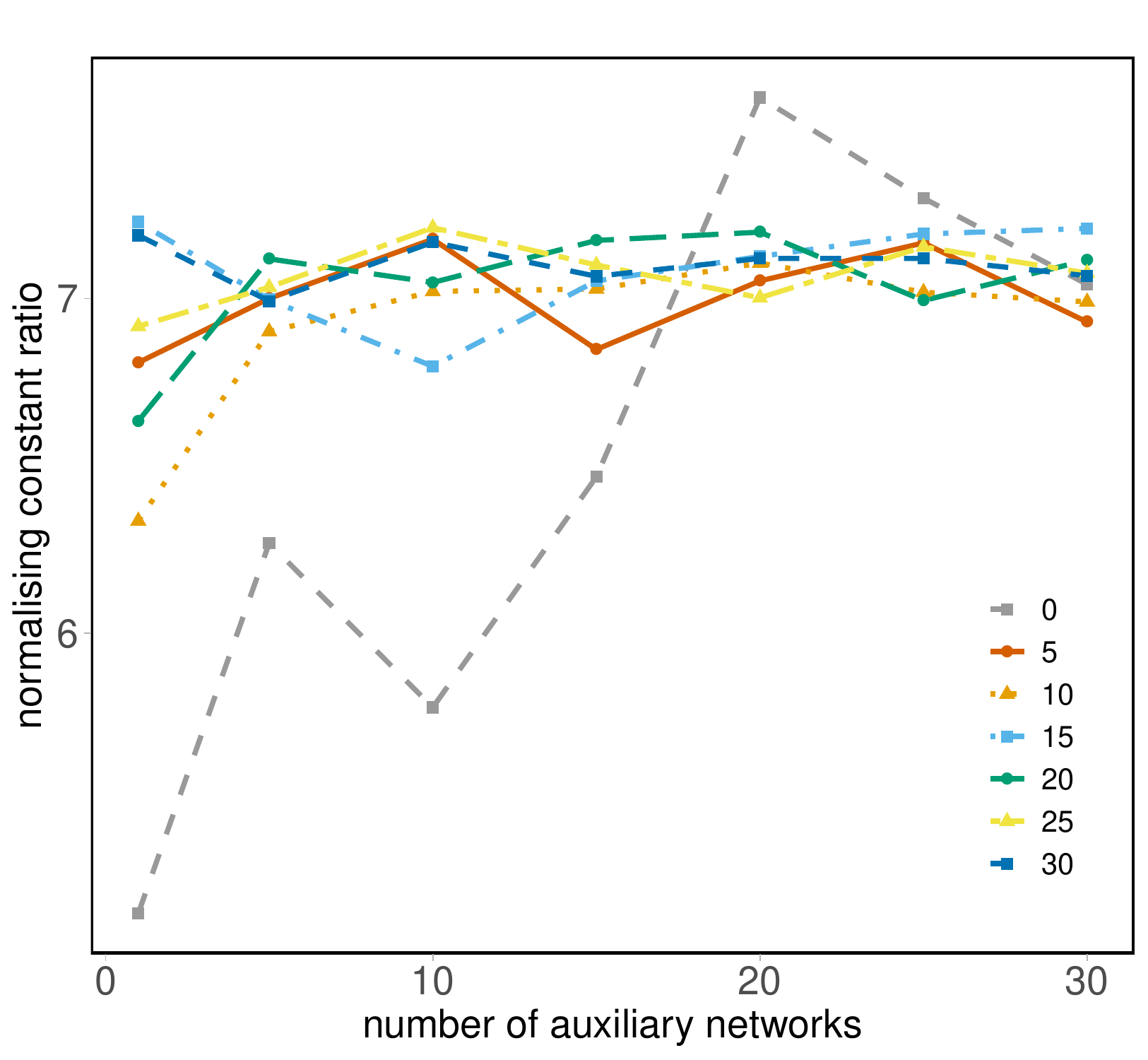}
		\includegraphics[width=4.5cm]{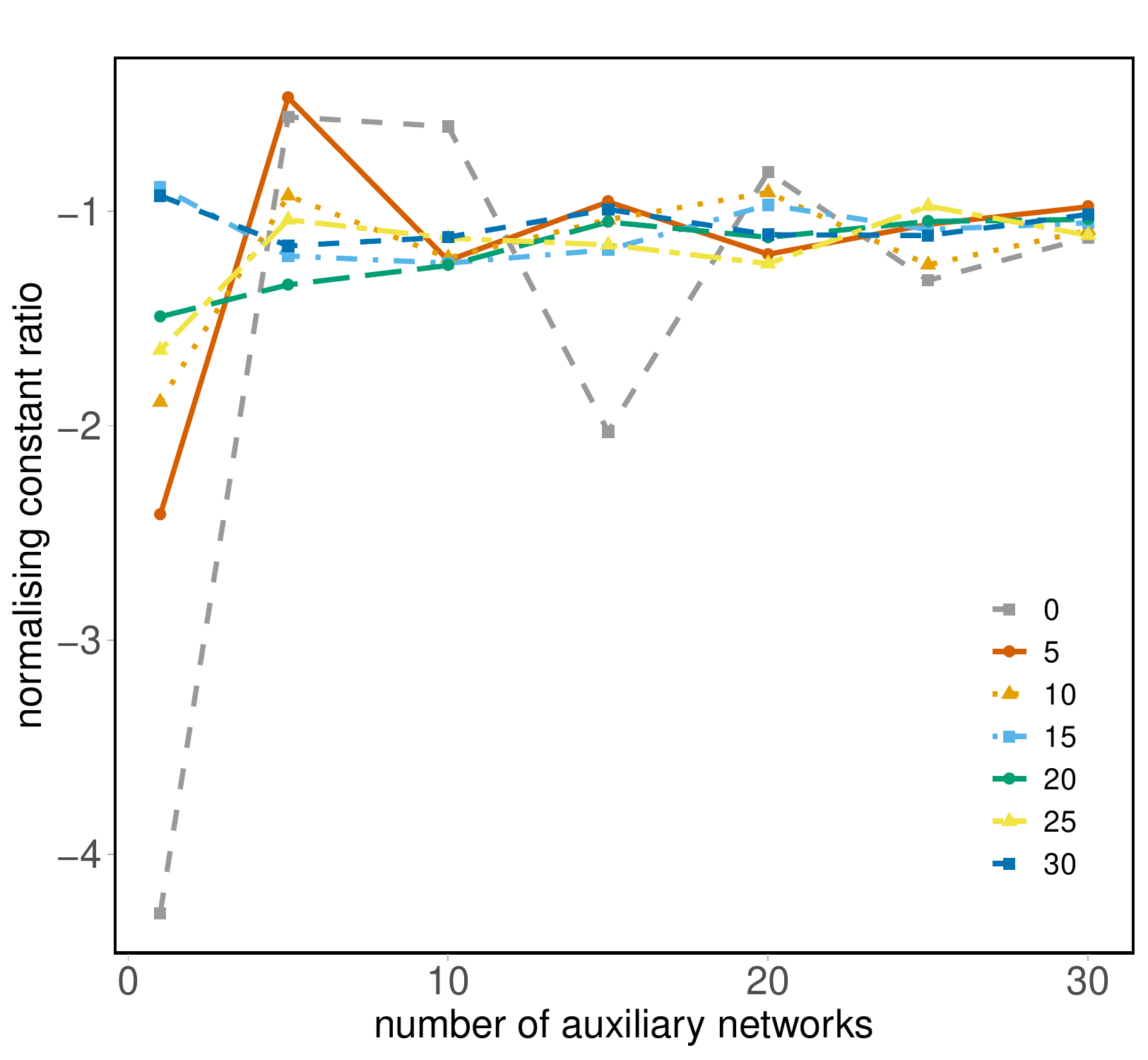}
		\includegraphics[width=4.5cm]{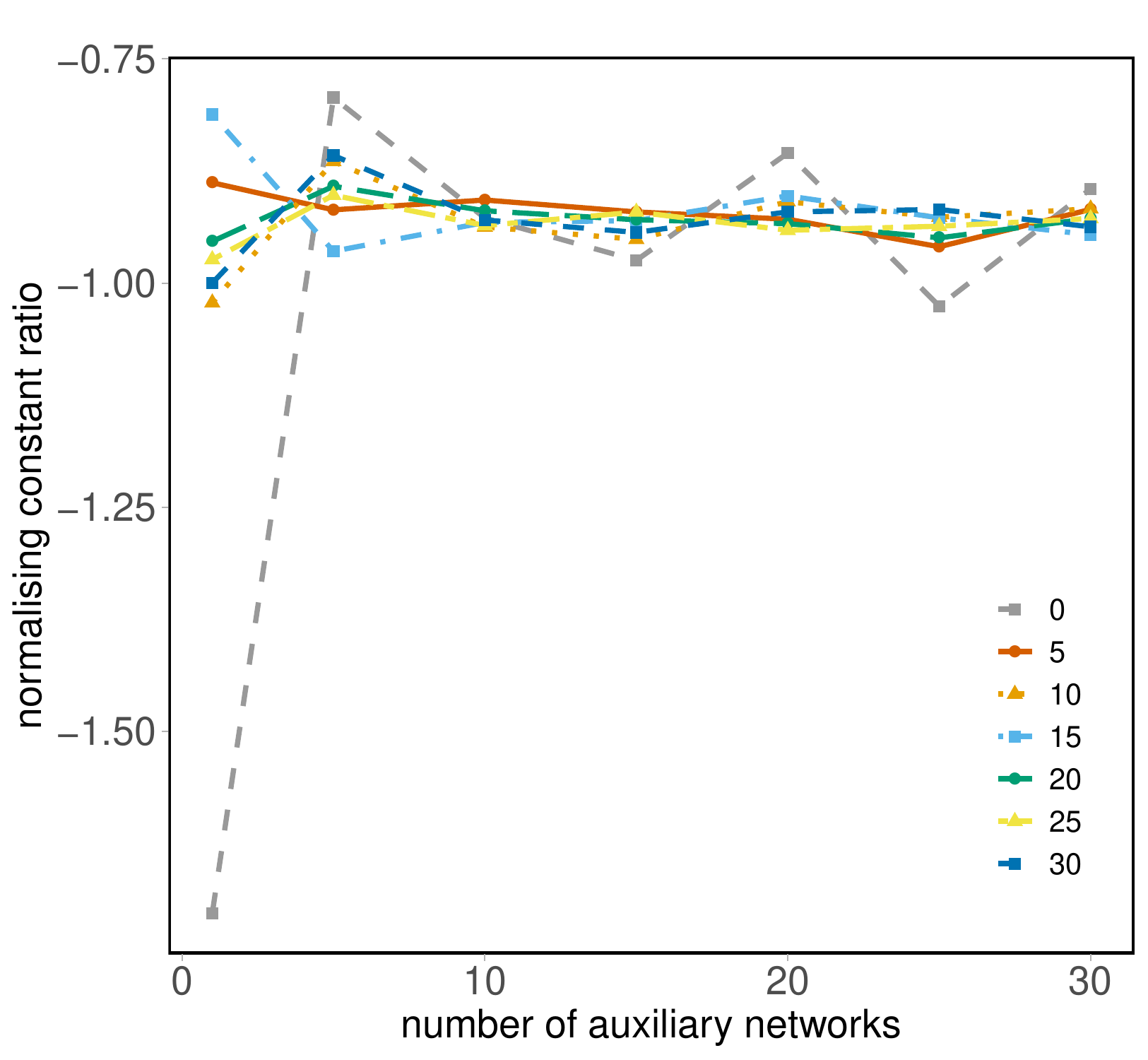}
	\end{center}
	\caption{Estimation of normalizing constant ratios when the compared parameters are close to each other. Three plots correspond to three repetitions of the parameters.}\label{nc2}
\end{figure}

In the second simulation, we show how $m_1, m_2$ affect the intermediate importance sampling estimation when the compared parameters are close. 
Here, we generate a sample $\theta_1$ from the prior $\mathcal{N} (\mu_0,\Sigma_0)$, and propose $\theta_2$ from a normal distribution $\mathcal{N}(\theta_1,\Sigma_p)$.
Then we estimate normalizing constant ratio $k(\theta_2)/k(\theta_1)$ with different $m_1, m_2$, and show the estimation in Figure \ref{nc2}.  
As we can see, all lines merge together with increasing $m_1$ and $m_2$, indicating that the intermediate importance sampling estimation is consistent. The simple importance sampling estimation (the line with $m_1=0$) has big variations, and intermediate importance sampling estimators (lines with $m_1>0$) are more stable. %The estimation line of $m_1=5$ is more stable. 

We recommend $m_1=2, m_2=10$ for MMCMH algorithm and $m_1=5, m_2=10$ for the posterior membership sampling as initial values, and similar techniques can be applied to choose $m_1, m_2$ in the specific dataset.  %All lines merge together with increasing $m_1$ and $m_2$, meaning that except for $m_1=0$ when there is no intermediate distributions used. %We choose $m_1=2,m_2=10$ in the MCMC step and $m_1=5,m_2=10$ in the allocation step. Here, we provide a way to initialize the values of $m_1, m_2$, which can be adjusted during iterations.
%\section*{Appendix 2: International Trade Networks}
\subsection*{1.2  International Trade Networks}
We also apply the proposed DPM-ERGMs to a world trade network ensemble. The ensemble of trade networks is observed on 60 countries ($n=60$) over the period 2001-2016 ($N=16$), denoted as ${y_i (i=1,2,\dots,16)}$. The networks are built based on the annual import data between every two countries from the UN Comtrade website \footnote{https://comtrade.un.org/}. The trade amount was collected in constant 2010 US dollars. A directed edge exists from node $r$ to $s$, $y_{rs,i}=1$, if the import amount from country $r$ to $s$ is more than 3 billion dollars at year $i$. %We choose 3 billion to be the threshold value. % because in this way the average density of the ensemble of networks is 0.15. Most real networks are sparse and 0.15 is a common number for network density.% Considering the completeness of the dataset, we choose 60 countries with 16-year period.
The geographic distance between countries, represented by a matrix $X$, is an important factor in analyzing trade relationship. Here, we treat distance as edge covariate and explore its influence on the trade ensemble. The distance between countries is calculated using the coordinate of the capital city, downloaded from CEPII database\footnote{http://www.cepii.fr}.

In this application, we choose four statistics to explore the ensembles of trade networks from different aspects,
\begin{itemize}
	\item $S^1(y_i)=\sum_{r\neq s}y_{rs,i}$, the total number of edges in network $y_i$. The density of trade networks can reflect the universality of global trade relationship.
	\item $S^2(y_i)=\sum_{r\neq s}y_{rs,i}y_{sr,i}$, the total number of mutual edges. The mutual edge in trade networks stands for bilateral trade. It is helping in understanding trade types.
	\item $S^3(y_i)=e^{\phi}\sum_{k=1}^{n-2}\{1-(1-e^{-\phi})^k\}EP_k(y_i)$, $\phi=0.25$, geometrically weighted edgewise shared partner, GWESP, a representation for transitivity. $EP_k(y_i)$ is the number of connected pairs that have $k$ common neighbors.
	\item $S^4(y_i)=\sum_{r\neq s}y_{rs,i}X_{rs}$, the effect of the distance covariate. This helps to explore how distance affects the trade network structure.
\end{itemize}
\begin{figure*}[!htbp]{}
	\centering
	%\subfigure[The number of clusters at every iteration]{\includegraphics[width=7cm]{Trade_full_z_100K}}
	%\subfigure[The frequency of allocating to each group after 50000 burn in]{\includegraphics[width=7cm]{Trade_full_z50K_freq}}
	\includegraphics[width=6.5cm]{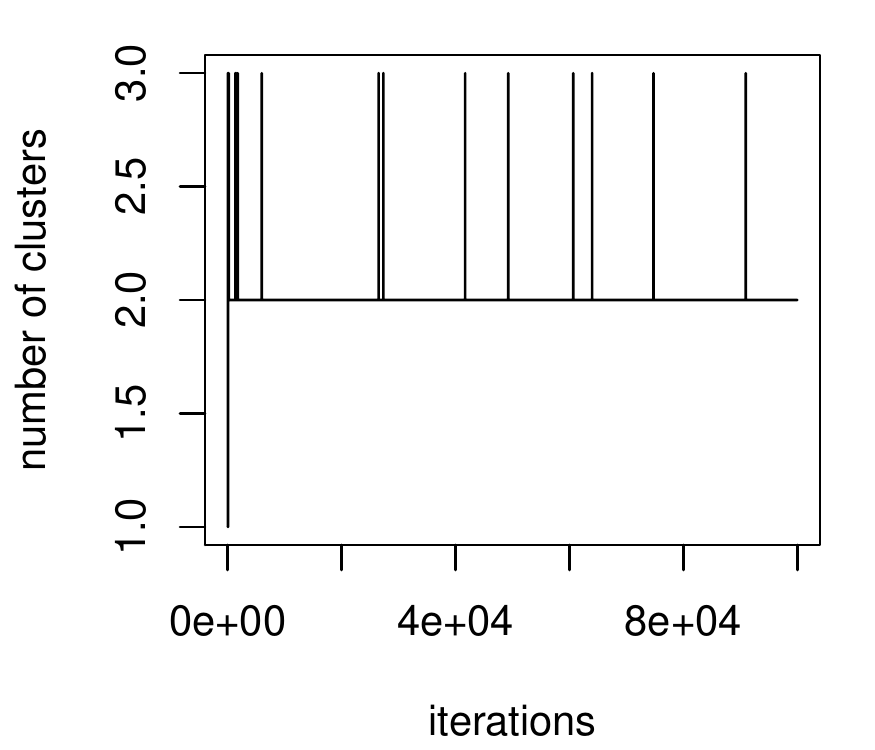}
	\includegraphics[width=6.5cm]{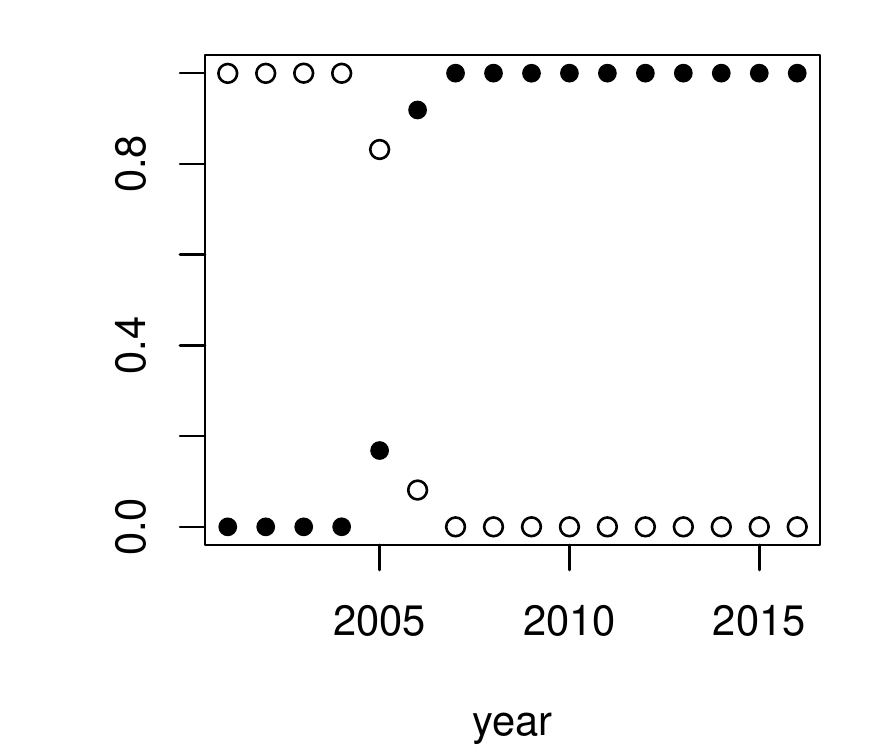}
	\caption{Clustering results of trade ensemble using IIMS sampling algorithm. Left: the number of clusters at every iteration. Right: the frequency of allocating to each group after 50,000 burn in.} \label{TCF}
\end{figure*}

We ran 100,000 iterations using IIMS sampling algorithm with the first 50,000 iterations as burn in. The hyperparameter and initial values are set as follows, $\theta_0=(-2,0,0,0)$ for ERGM parameter initial, $\mu_0=(-3,0,0,0)$, $\Sigma_0=4^2I_4$ for ERGM parameter prior, a diagonal matrix $\Sigma_q$ with diagonal entries $(0.05^2,0.02^2,0.02^2,0.02^2)$ for the variance of the proposal distribution in MMCMH, $\text{Beta} (1,0.1)$ for the sticking breaking prior. In intermediate importance sampling, we choose $m_1=2,m_2=10$ for MMCMH and $m_1=5,m_2=10$ for posterior membership sampling. As shown in Figure \ref{TCF}, the ensemble of trade networks is clustered into 2 groups. Group 1 corresponds to networks of earlier years, from 2001 to 2005, and group 2 is formed by networks of later years, between 2006 and 2016. The acceptance ratio is 0.45, 0.30 for two groups separately. The network membership is closely related to the time, which is reasonable as trade networks are collected over time. %The clustering results are reasonale because the ensemble of trade networks is time dependent. %The period 2001-2016 can be divided into two phases with different characteristics.
%The trade size is growing gradually with time. The first 5 networks belong to the first stage, where the trade size is smaller. The rest of 11 networks comprise the second stage, where the interactions among countries are stronger.
\begin{figure*}[!htbp]{}
	\centering
	\text{IIMS}
	\includegraphics[width=16cm]{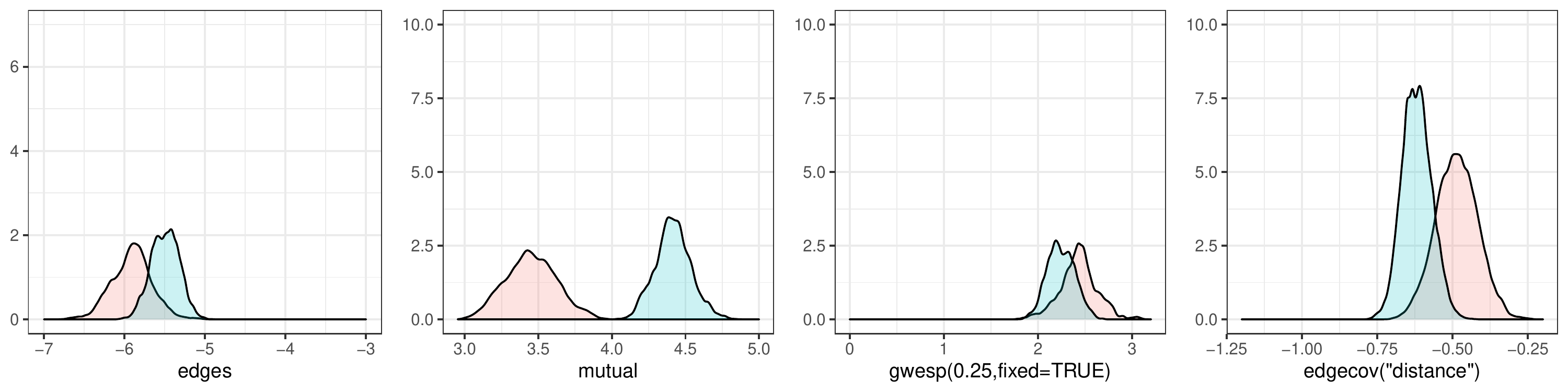}\\
	\text{PMS}
	\includegraphics[width=16cm]{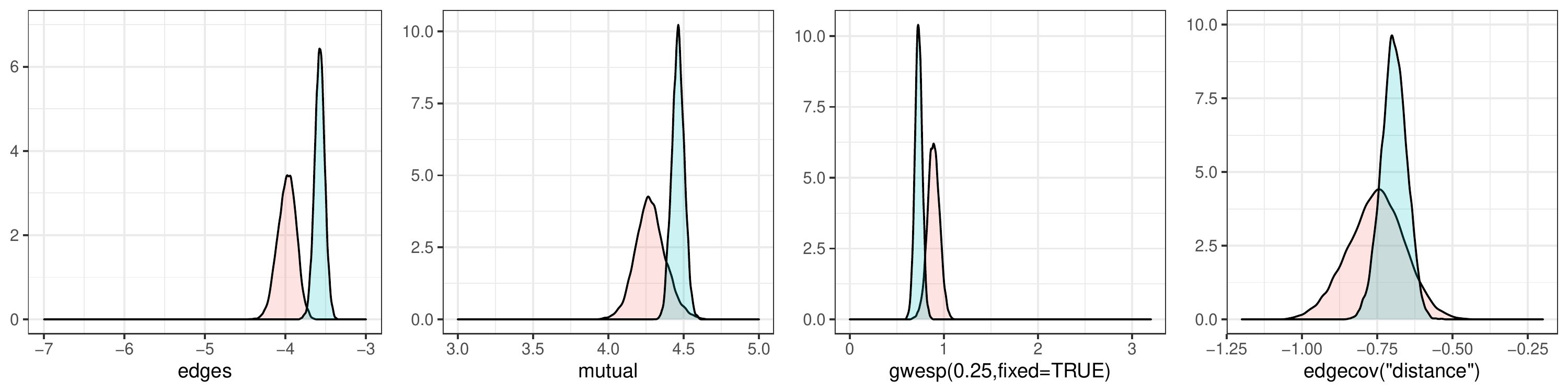}\\
	%\text{PMG}
	%\includegraphics[width=18cm]{Trade_Yin_50Kburnin_density2in1_14}
	\caption{Density plots of estimation for groups 1 (red), 2 (blue) in trade ensemble. %The first 50000 samples are used as burn in. Red colour is for group 1 while the blue is for group 2. Top: infinite full likelihood method; middle: infinite pseudo likelihood method; bottom: finite pseudo likelihood. 
	}\label{TD}
\end{figure*}

The characteristics of each group can be further described using a group-specific ERGM and the comparisons between groups can be performed by comparing the parameters of each ERGM.
The density plots for the posterior samples are shown in the first row of Figure \ref{TD}.
%We first compare the estimation between groups.  
As we can see, group 2 has bigger density parameter than group 1, meaning that the trade relationships are denser. It also has bigger mutuality, which indicates that bilateral trade is more common. More countries prefer to form a mutual trade relationship with their trading partners. The smaller transitivity coefficient of group 2 suggests that the international trade is becoming more universal, although that the local clustering phenomenon still exists, implied by the positive transitivity parameter. %, which is because the transitivity coefficient is positive.
\begin{figure}{}
	\centering
	\includegraphics[width=6.5cm]{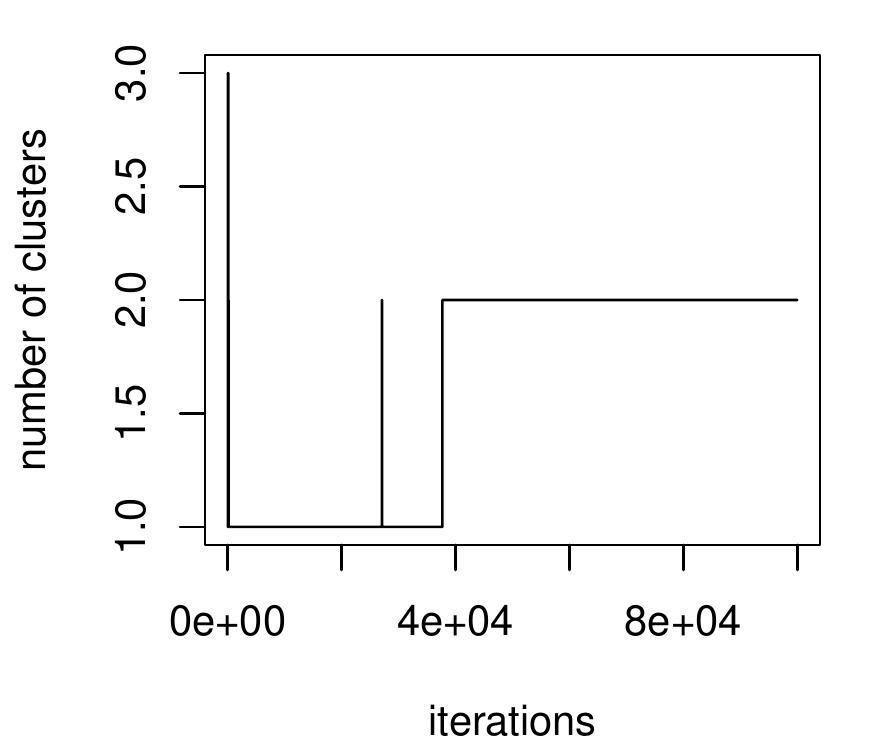}
	\includegraphics[width=6.5cm]{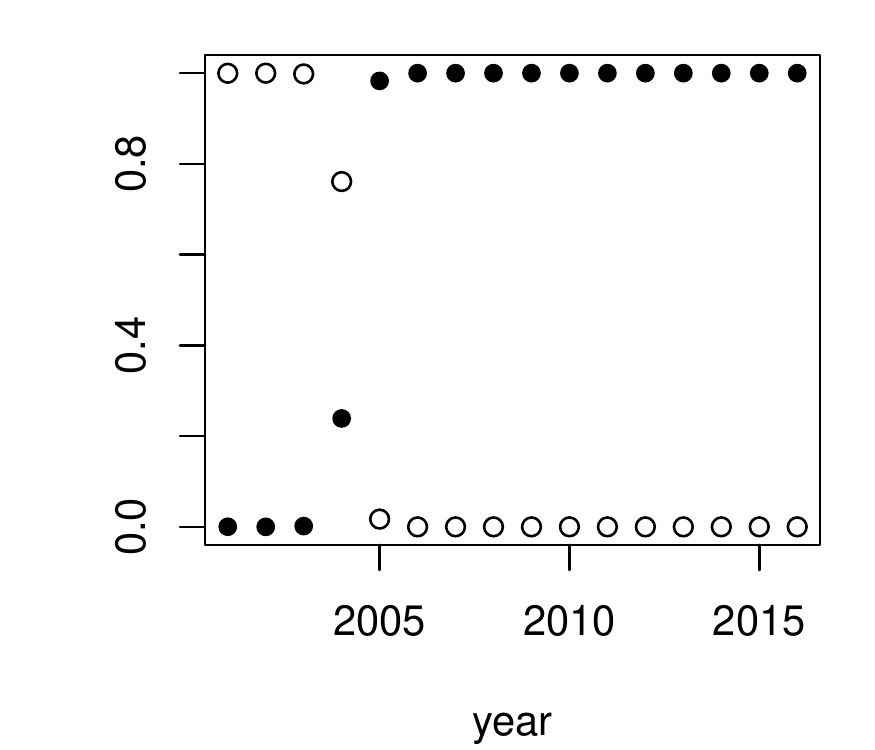}
	\caption{Clustering results of trade ensemble using PMS sampling algorithm. Left: the number of clusters at every iteration. Right: the frequency of allocating to each group after 50,000 burn in.} \label{TCP}
\end{figure}

Next, we ran the PMS sampling algorithm 100,000 iterations. The clustering result is displayed in Figure \ref{TCP}, which is similar to the IIMS algorithm. Networks from 2001 to 2004 are in the group 1 and networks between 2005 and 2016 are in the group 2. The acceptance ratio of each group is 0.59, 0.32. The density plots for each group are shown in the second row of Figure \ref{TD}. Regardless of the similar clustering result, the density plots for ERGM parameter estimation are quite different. Comparing with the IIMS algorithm, PMS provides a narrow and sharp estimation, because pseudo likelihood method underestimates the variance of estimation.
Moreover, the coefficient for GWESP term from PMS method is much smaller compared with the IIMS method. This is because the pseudo likelihood method can not capture the dependent structures within a network.
\begin{figure*}[!htbp]{}
	\centering
	\text{IIMS}
	\includegraphics[width=16cm]{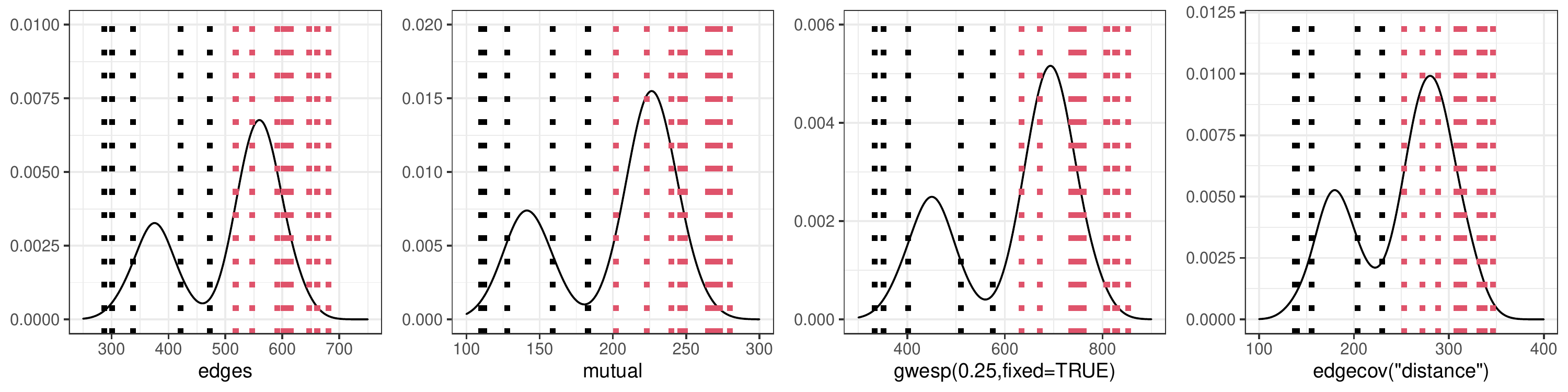}\\
	\text{PMS}
	\includegraphics[width=16cm]{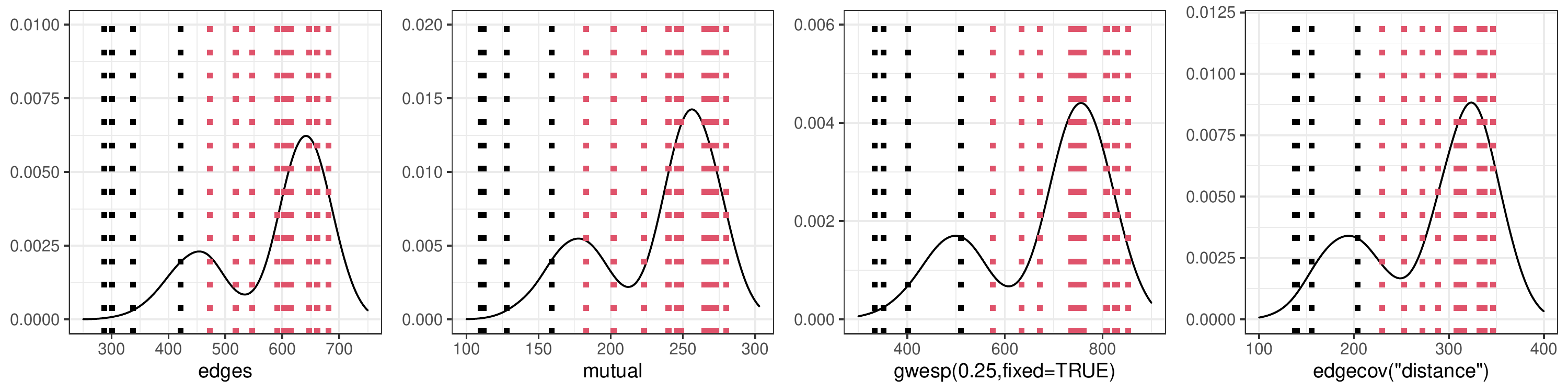}
	\caption{Density plots of network statistics, edges, mutual, GWESP and edgecov, based on networks simulated from estimation. The black vertical lines represent networks of group 1 and the red lines stand for networks of group 2.} \label{TA}
\end{figure*}

%The modes of mutual and distance terms are similar for group 2.
%It is possible to have one more paragrah to tell the story of trade changes, like more dense, more mutual but less local clsuterign effect, more global. Distance is affecting less.
In order to assess the model results, we generate 1000 networks from the estimated model, and plot the simulated network statistics in Figure \ref{TA}. The black dots stand for networks of group 1 and red dots represent networks of group 2. In the first row of the figure, the density plots of simulated network statistics are close to the observed samples, indicating that IIMS algorithm provides good estimation to the data. However, in the second row, the posterior mode of the first group is far away from the 4 samples in the group, implying that the model is not a good fit to the data. As we mentioned before, networks in group 1 have strong transitivity, which can not be captured by PMS algorithm. %Both full and pseudo likelihood methods behave well for the second group.

Furthermore, we calculate the distance between observed network statistics and simulated network statistics.
Results are shown in Table \ref{Tdis}. IIMS method fits the networks in group 1 better than PMS method as the distance 89344 is much smaller than 145566. 
\begin{table}[h!]
	\caption{The distance between observed network statistics and simulated network statistics.} \label{Tdis}
	\centering
	\begin{tabular}{ccc} 
		Method & Group 1  & Group 2 \\\hline
		IIMS & 80344 & 188624\\ 
		PMS & 145566 & 175461 \\ 
	\end{tabular}
\end{table}
\begin{figure}{}
	\centering
	\includegraphics[width=6.5cm]{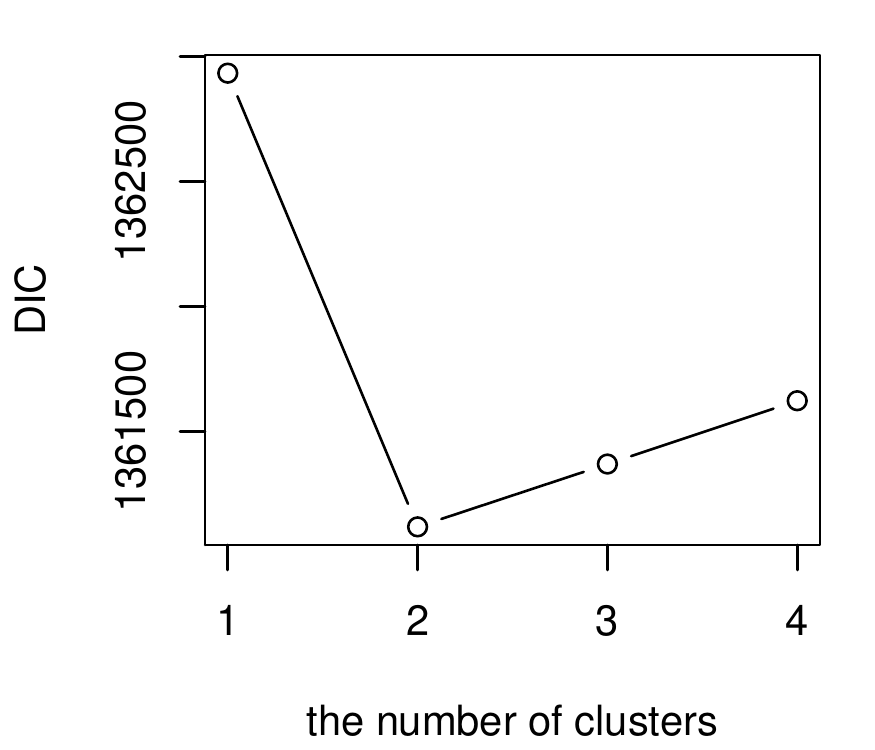}
	\includegraphics[width=6.5cm]{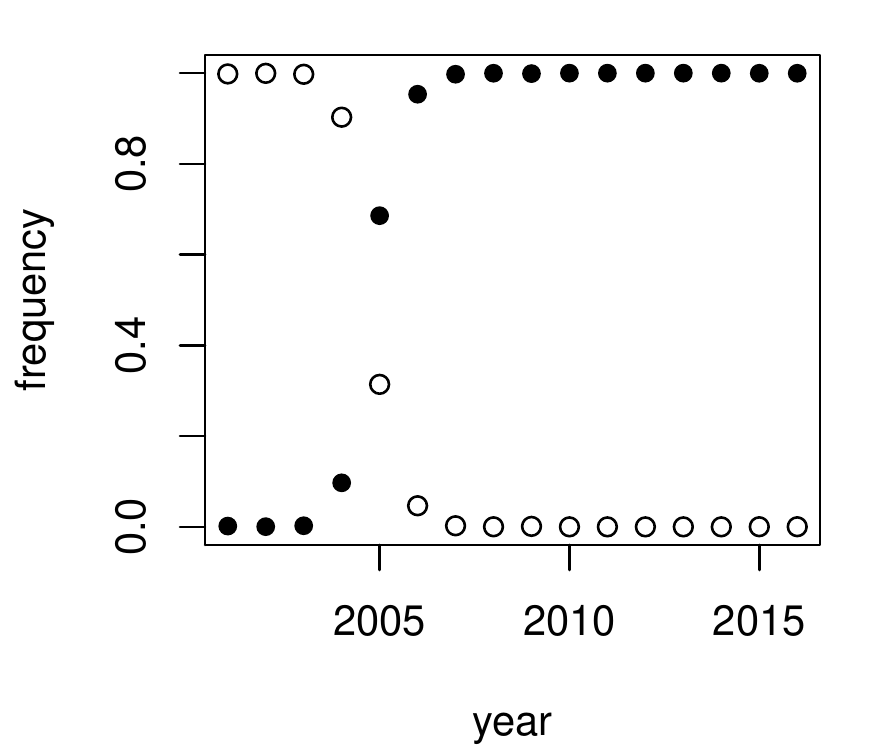}
	\caption{Clustering results of trade ensemble using PMG sampling algorithm. Left: DIC values for different number of clusters. Right: the frequency of allocating to each group after 50,000 burn in when there are 2 groups.} \label{TCY}
\end{figure}

For comparison, we also applied the pseudo likelihood Metropolis-within-Gibbs (PMG) sampling algorithm, developed by \cite{Yin2020} for a finite mixture of ERGMs. The prior for the group parameter is the same as it for the infinite method. %For the mixing proportion, we use a Dirichlet distribution as prior. 
Without knowing the number of clusters in advance, we fit the model with the number of clusters $K=1,2,3,4$ in sequence and calculate deviance information criteria (DIC) for each model accordingly.
The DIC value with different number of clusters is shown in Figure \ref{TCY}. The best model is the one with the smallest DIC value, meaning that the number of clusters is chosen to be 2 here. Networks between 2001 and 2004 are allocated to the first group and networks between 2005 and 2016 are in the second group. The individual network membership and the posterior density of each group from PMG algorithm are the same as our proposed PMS algorithm.

In this simulation, IIMS algorithm clustered the trade ensemble into 2 groups and fit each group with a different ERGM. PMS algorithm provides similar clustering result to IIMS, but the networks in group 1 are fit poorly because the pseudo likelihood method failed to capture the transitivity of trade networks. The results of PMG algorithm are comparable with PMS algorithm, which guaranteed our method.
%In this simulation, both the proposed IIMS and PMS sampling algorithms clustered the trade ensemble into 2 groups. The IIMS algorithm provides a better posterior samples as it performs full Bayesian inference for the estimation. The results are confirmed by PMG algorithm.
%Of all the 5000 iterations, network 2005 is clustered into group 2 for 562 times, network 2006 is grouped into group 1 for 700 times, and network 2009 is classified into group 1 for 2 times. Networks of years 2005, 2006, 2009 are shown in black, dark blue and light blue lines individually in Figure \ref{TA}. It can be seen clearly from the plot that networks of years 2005 and 2006 are located between two density plots, which agrees with the fact that they are the transitions between the two groups. We also notice that in the first group, networks are growing with time order which is shown in the plot that the vertical lines standing for 2001-2005 are locating from left to right. However, for network 2009, it is closer to network 2006 than the rest of networks in group 2 (green vertical line), including its neighbours 2008 and 2009. This agrees with the fact that the financial crisis 2008 decreased the trade relationship significantly. The green lines, standing for networks of years 2007, 2008, 2010-2016, are closer to each other than the other lines, which is consistent with the fact that the economic grew slower than previous years during this period.

\end{document}